%% file: paper.tex
\documentclass[letterpaper,twocolumn,10pt]{article}
\input{macros}

\AtBeginDocument{%
  }

\begin{document}
\title{\bf Rex: Safe and usable kernel extensions in Rust}
\title{\bf Safe and usable kernel extensions with Rex}

\author{
  Jinghao Jia$^{*}$, Ruowen Qin$^{*}$, Milo Craun$^{\dagger}$, Egor Lukiyanov$^{\dagger}$, Ayush Bansal$^{*}$,\\[3pt]
  Michael V. Le$^{\ddagger}$, Hubertus Franke$^{\ddagger}$, Hani Jamjoom$^{\ddagger}$, Tianyin Xu$^{*}$, Dan Williams$^{\dagger}$\\[7.5pt]
  $^{*}$University of Illinois Urbana-Champaign\ \ \ \  $^{\dagger}$Virginia Tech\ \ \ \  $^{\ddagger}$IBM Research\\[17.5pt]
}





\maketitle

\input{abstract}

\input{intro}
\input{background}
\input{motivation}
\input{safety_model}

\input{principle}
\input{impl}
\input{eval}

\input{discussion}
\input{related}
\input{conclusion}
{\small
\bibliographystyle{acm}
\bibliography{ref}
}


\end{document}

%% file: macros.tex
\PassOptionsToPackage{hyphens}{url}
\usepackage{usenix2021}
\usepackage{graphicx}
\usepackage{fancybox}
\usepackage{listings}
\usepackage{xcolor}
\usepackage{subcaption}
\usepackage{setspace}
\usepackage{enumitem}
\usepackage{booktabs}
\usepackage{comment}

\usepackage{titling}
\postauthor{\end{tabular}\vspace{-0.5in}\end{center}}
\date{}
\usepackage{titlesec}
\titleformat{\subsection}
  {\normalfont\fontsize{11}{10}\bfseries}{\thesubsection}{1em}{}

\definecolor{ballblue}{rgb}{0.13, 0.67, 0.8}
\definecolor{bondiblue}{rgb}{0.0, 0.58, 0.71}
\definecolor{cobalt}{rgb}{0.0, 0.28, 0.67}
\definecolor{copper}{rgb}{0.72, 0.45, 0.2}
\definecolor{darkblue}{rgb}{0.0, 0.0, 0.55}
\definecolor{darkspringgreen}{rgb}{0.09, 0.45, 0.27}
\definecolor{gray}{gray}{0.5}
\definecolor{LightGray}{gray}{0.85}
\definecolor{VeryLightGray}{gray}{0.90}
\definecolor{ForestGreen}{rgb}{0.13, 0.55, 0.13}
\definecolor{Maroon}{rgb}{0.5, 0.0, 0.0}

\newcommand{\para}[1]{\smallskip\noindent {\bf #1} }
\newcommand{\comments}[1]{}

\iffalse   
\newcommand{\tianyin}[1]{{{\color{red} [ty: #1]}}}
\newcommand{\jinghao}[1]{}
\newcommand{\new}[1]{#1}
\newcommand{\juowen}[1]{{\color{gray} Ruowen: #1}}
\newcommand{\djw}[1]{{\color{red}{\em (djw: #1)}}}
\newcommand{\raj}[1]{{\color{red}{\em (raj: #1)}}}
\newcommand{\mvle}[1]{{\color{darkblue}{\em (mvle: #1)}}}
\newcommand{\milo}[1]{{\color{orange}{\em (milo: #1)}}}
\newcommand{\hubertus}[1]{{\color{orange}{\em (frankeh: #1)}}}
\newcommand{\ayushb}[1]{{\color{ForestGreen} (ayushb: #1)}}

\else
\newcommand{\tianyin}[1]{}
\newcommand{\jinghao}[1]{}
\newcommand{\new}[1]{#1}
\newcommand{\juowen}[1]{}
\newcommand{\djw}[1]{}
\newcommand{\raj}[1]{}
\newcommand{\mvle}[1]{}
\newcommand{\milo}[1]{}
\newcommand{\hubertus}[1]{}
\newcommand{\ayushb}[1]{}

\fi

\newcommand{\projname}{Rex}
\newcommand{\gap}{language-verifier gap}

\lstdefinelanguage{Rust}{%
  sensitive%
, morecomment=[l]{//}%
, morecomment=[s]{/*}{*/}%
, moredelim=[s][{\itshape\color[rgb]{0,0,0.75}}]{\#[}{]}%
, morestring=[b]{"}%
, alsodigit={}%
, alsoother={}%
, alsoletter={!}%
, morekeywords={break, continue, else, for, if, in, loop, match, return, while}  
, morekeywords={as, const, let, move, mut, ref, static}  
, morekeywords={dyn, enum, fn, impl, Self, self, struct, trait, type, union, use, where}  
, morekeywords={crate, extern, mod, pub, super}  
, morekeywords={unsafe}  
, morekeywords={abstract, alignof, become, box, do, final, macro, offsetof, override, priv, proc, pure, sizeof, typeof, unsized, virtual, yield}  
, morekeywords=[2]{Add, AddAssign, Any, AsciiExt, AsInner, AsInnerMut, AsMut, AsRawFd, AsRawHandle, AsRawSocket, AsRef, Binary, BitAnd, BitAndAssign, Bitor, BitOr, BitOrAssign, BitXor, BitXorAssign, Borrow, BorrowMut, Boxed, BoxPlace, BufRead, BuildHasher, CastInto, CharExt, Clone, CoerceUnsized, CommandExt, Copy, Debug, DecodableFloat, Default, Deref, DerefMut, DirBuilderExt, DirEntryExt, Display, Div, DivAssign, DoubleEndedIterator, DoubleEndedSearcher, Drop, EnvKey, Eq, Error, ExactSizeIterator, ExitStatusExt, Extend, FileExt, FileTypeExt, Float, Fn, FnBox, FnMut, FnOnce, Freeze, From, FromInner, FromIterator, FromRawFd, FromRawHandle, FromRawSocket, FromStr, FullOps, FusedIterator, Generator, Hash, Hasher, Index, IndexMut, InPlace, Int, Into, IntoCow, IntoInner, IntoIterator, IntoRawFd, IntoRawHandle, IntoRawSocket, IsMinusOne, IsZero, Iterator, JoinHandleExt, LargeInt, LowerExp, LowerHex, MetadataExt, Mul, MulAssign, Neg, Not, Octal, OpenOptionsExt, Ord, OsStrExt, OsStringExt, Packet, PartialEq, PartialOrd, Pattern, PermissionsExt, Place, Placer, Pointer, Product, Put, RangeArgument, RawFloat, Read, Rem, RemAssign, Seek, Shl, ShlAssign, Shr, ShrAssign, Sized, SliceConcatExt, SliceExt, SliceIndex, Stats, Step, StrExt, Sub, SubAssign, Sum, Sync, TDynBenchFn, Terminal, Termination, ToOwned, ToSocketAddrs, ToString, Try, TryFrom, TryInto, UnicodeStr, Unsize, UpperExp, UpperHex, WideInt, Write}
, morekeywords=[2]{Send}  
, morekeywords=[3]{bool, char, f32, f64, i8, i16, i32, i64, isize, str, u8, u16, u32, u64, unit, usize, i128, u128}  
, morekeywords=[4]{Err, false, None, Ok, Some, true}  
, morekeywords=[3]{AccessError, Adddf3, AddI128, AddoI128, AddoU128, ADDRESS, ADDRESS64, addrinfo, ADDRINFOA, AddrParseError, Addsf3, AddU128, advice, aiocb, Alignment, AllocErr, AnonPipe, Answer, Arc, Args, ArgsInnerDebug, ArgsOs, Argument, Arguments, ArgumentV1, Ashldi3, Ashlti3, Ashrdi3, Ashrti3, AssertParamIsClone, AssertParamIsCopy, AssertParamIsEq, AssertUnwindSafe, AtomicBool, AtomicPtr, Attr, auxtype, auxv, BackPlace, BacktraceContext, Barrier, BarrierWaitResult, Bencher, BenchMode, BenchSamples, BinaryHeap, BinaryHeapPlace, blkcnt, blkcnt64, blksize, BOOL, boolean, BOOLEAN, BoolTrie, BorrowError, BorrowMutError, Bound, Box, bpf, BTreeMap, BTreeSet, Bucket, BucketState, Buf, BufReader, BufWriter, Builder, BuildHasherDefault, BY, BYTE, Bytes, CannotReallocInPlace, cc, Cell, Chain, CHAR, CharIndices, CharPredicateSearcher, Chars, CharSearcher, CharsError, CharSliceSearcher, CharTryFromError, Child, ChildPipes, ChildStderr, ChildStdin, ChildStdio, ChildStdout, Chunks, ChunksMut, ciovec, clock, clockid, Cloned, cmsgcred, cmsghdr, CodePoint, Color, ColorConfig, Command, CommandEnv, Component, Components, CONDITION, condvar, Condvar, CONSOLE, CONTEXT, Count, Cow, cpu, CRITICAL, CStr, CString, CStringArray, Cursor, Cycle, CycleIter, daddr, DebugList, DebugMap, DebugSet, DebugStruct, DebugTuple, Decimal, Decoded, DecodeUtf16, DecodeUtf16Error, DecodeUtf8, DefaultEnvKey, DefaultHasher, dev, device, Difference, Digit32, DIR, DirBuilder, dircookie, dirent, dirent64, DirEntry, Discriminant, DISPATCHER, Display, Divdf3, Divdi3, Divmoddi4, Divmodsi4, Divsf3, Divsi3, Divti3, dl, Dl, Dlmalloc, Dns, DnsAnswer, DnsQuery, dqblk, Drain, DrainFilter, Dtor, Duration, DwarfReader, DWORD, DWORDLONG, DynamicLibrary, Edge, EHAction, EHContext, Elf32, Elf64, Empty, EmptyBucket, EncodeUtf16, EncodeWide, Entry, EntryPlace, Enumerate, Env, epoll, errno, Error, ErrorKind, EscapeDebug, EscapeDefault, EscapeUnicode, event, Event, eventrwflags, eventtype, ExactChunks, ExactChunksMut, EXCEPTION, Excess, ExchangeHeapSingleton, exit, exitcode, ExitStatus, Failure, fd, fdflags, fdsflags, fdstat, ff, fflags, File, FILE, FileAttr, filedelta, FileDesc, FilePermissions, filesize, filestat, FILETIME, filetype, FileType, Filter, FilterMap, Fixdfdi, Fixdfsi, Fixdfti, Fixsfdi, Fixsfsi, Fixsfti, Fixunsdfdi, Fixunsdfsi, Fixunsdfti, Fixunssfdi, Fixunssfsi, Fixunssfti, Flag, FlatMap, Floatdidf, FLOATING, Floatsidf, Floatsisf, Floattidf, Floattisf, Floatundidf, Floatunsidf, Floatunsisf, Floatuntidf, Floatuntisf, flock, ForceResult, FormatSpec, Formatted, Formatter, Fp, FpCategory, fpos, fpos64, fpreg, fpregset, FPUControlWord, Frame, FromBytesWithNulError, FromUtf16Error, FromUtf8Error, FrontPlace, fsblkcnt, fsfilcnt, fsflags, fsid, fstore, fsword, FullBucket, FullBucketMut, FullDecoded, Fuse, GapThenFull, GeneratorState, gid, glob, glob64, GlobalDlmalloc, greg, group, GROUP, Guard, GUID, Handle, HANDLE, Handler, HashMap, HashSet, Heap, HINSTANCE, HMODULE, hostent, HRESULT, id, idtype, if, ifaddrs, IMAGEHLP, Immut, in, in6, Incoming, Infallible, Initializer, ino, ino64, inode, input, InsertResult, Inspect, Instant, int16, int32, int64, int8, integer, IntermediateBox, Internal, Intersection, intmax, IntoInnerError, IntoIter, IntoStringError, intptr, InvalidSequence, iovec, ip, IpAddr, ipc, Ipv4Addr, ipv6, Ipv6Addr, Ipv6MulticastScope, Iter, IterMut, itimerspec, itimerval, jail, JoinHandle, JoinPathsError, KDHELP64, kevent, kevent64, KV, l4, LARGE, lastlog, launchpad, Layout, Lazy, lconv, Leaf, LeafOrInternal, Lines, LinesAny, LineWriter, linger, linkcount, LinkedList, load, locale, LocalKey, LocalKeyState, Location, lock, LockResult, loff, LONG, lookup, lookupflags, LookupHost, LPBOOL, LPBY, LPBYTE, LPCSTR, LPCVOID, LPCWSTR, LPDWORD, LPFILETIME, LPHANDLE, LPOVERLAPPED, LPPROCESS, LPPROGRESS, LPSECURITY, LPSTARTUPINFO, LPSTR, LPVOID, LPWCH, LPWIN32, LPWSADATA, LPWSAPROTOCOL, LPWSTR, Lshrdi3, Lshrti3, lwpid, M128A, mach, major, Map, mcontext, Metadata, Metric, MetricMap, mflags, minor, mmsghdr, Moddi3, mode, Modsi3, Modti3, MonitorMsg, MOUNT, mprot, mq, mqd, msflags, msghdr, msginfo, msglen, msgqnum, msqid, Muldf3, Mulodi4, Mulosi4, Muloti4, Mulsf3, Multi3, Mut, Mutex, MutexGuard, MyCollection, n16, NamePadding, NativeLibBoilerplate, nfds, nl, nlink, NodeRef, NoneError, NonNull, NonZero, nthreads, NulError, OccupiedEntry, off, off64, oflags, Once, OnceState, OpenOptions, Option, Options, OptRes, Ordering, OsStr, OsString, Output, OVERLAPPED, Owned, Packet, PanicInfo, Param, ParseBoolError, ParseCharError, ParseError, ParseFloatError, ParseIntError, ParseResult, Part, passwd, Path, PathBuf, PCONDITION, PCONSOLE, Peekable, PeekMut, Permissions, PhantomData, pid, Pipes, PlaceBack, PlaceFront, PLARGE, PoisonError, pollfd, PopResult, port, Position, Powidf2, Powisf2, Prefix, PrefixComponent, PrintFormat, proc, Process, PROCESS, processentry, protoent, PSRWLOCK, pthread, ptr, ptrdiff, PVECTORED, Queue, radvisory, RandomState, Range, RangeFrom, RangeFull, RangeInclusive, RangeMut, RangeTo, RangeToInclusive, RawBucket, RawFd, RawHandle, RawPthread, RawSocket, RawTable, RawVec, Rc, ReadDir, Receiver, recv, RecvError, RecvTimeoutError, ReentrantMutex, ReentrantMutexGuard, Ref, RefCell, RefMut, REPARSE, Repeat, Result, Rev, Reverse, riflags, rights, rlim, rlim64, rlimit, rlimit64, roflags, Root, RSplit, RSplitMut, RSplitN, RSplitNMut, RUNTIME, rusage, RwLock, RWLock, RwLockReadGuard, RwLockWriteGuard, sa, SafeHash, Scan, sched, scope, sdflags, SearchResult, SearchStep, SECURITY, SeekFrom, segment, Select, SelectionResult, sem, sembuf, send, Sender, SendError, servent, sf, Shared, shmatt, shmid, ShortReader, ShouldPanic, Shutdown, siflags, sigaction, SigAction, sigevent, sighandler, siginfo, Sign, signal, signalfd, SignalToken, sigset, sigval, Sink, SipHasher, SipHasher13, SipHasher24, Skip, SkipWhile, Slice, SmallBoolTrie, sockaddr, SOCKADDR, sockcred, Socket, SOCKET, SocketAddr, SocketAddrV4, SocketAddrV6, socklen, speed, Splice, Split, SplitMut, SplitN, SplitNMut, SplitPaths, SplitWhitespace, spwd, SRWLOCK, ssize, stack, STACKFRAME64, StartResult, STARTUPINFO, stat, Stat, stat64, statfs, statfs64, StaticKey, statvfs, StatVfs, statvfs64, Stderr, StderrLock, StderrTerminal, Stdin, StdinLock, Stdio, StdioPipes, Stdout, StdoutLock, StdoutTerminal, StepBy, String, StripPrefixError, StrSearcher, subclockflags, Subdf3, SubI128, SuboI128, SuboU128, subrwflags, subscription, Subsf3, SubU128, Summary, suseconds, SYMBOL, SYMBOLIC, SymmetricDifference, SyncSender, sysinfo, System, SystemTime, SystemTimeError, Take, TakeWhile, tcb, tcflag, TcpListener, TcpStream, TempDir, TermInfo, TerminfoTerminal, termios, termios2, TestDesc, TestDescAndFn, TestEvent, TestFn, TestName, TestOpts, TestResult, Thread, threadattr, threadentry, ThreadId, tid, time, time64, timespec, TimeSpec, timestamp, timeval, timeval32, timezone, tm, tms, ToLowercase, ToUppercase, TraitObject, TryFromIntError, TryFromSliceError, TryIter, TryLockError, TryLockResult, TryRecvError, TrySendError, TypeId, U64x2, ucontext, ucred, Udivdi3, Udivmoddi4, Udivmodsi4, Udivmodti4, Udivsi3, Udivti3, UdpSocket, uid, UINT, uint16, uint32, uint64, uint8, uintmax, uintptr, ulflags, ULONG, ULONGLONG, Umoddi3, Umodsi3, Umodti3, UnicodeVersion, Union, Unique, UnixDatagram, UnixListener, UnixStream, Unpacked, UnsafeCell, UNWIND, UpgradeResult, useconds, user, userdata, USHORT, Utf16Encoder, Utf8Error, Utf8Lossy, Utf8LossyChunk, Utf8LossyChunksIter, utimbuf, utmp, utmpx, utsname, uuid, VacantEntry, Values, ValuesMut, VarError, Variables, Vars, VarsOs, Vec, VecDeque, vm, Void, WaitTimeoutResult, WaitToken, wchar, WCHAR, Weak, whence, WIN32, WinConsole, Windows, WindowsEnvKey, winsize, WORD, Wrapping, wrlen, WSADATA, WSAPROTOCOL, WSAPROTOCOLCHAIN, Wtf8, Wtf8Buf, Wtf8CodePoints, xsw, xucred, Zip, zx}
, morekeywords=[5]{assert!, assert_eq!, assert_ne!, cfg!, column!, compile_error!, concat!, concat_idents!, debug_assert!, debug_assert_eq!, debug_assert_ne!, env!, eprint!, eprintln!, file!, format!, format_args!, include!, include_bytes!, include_str!, line!, module_path!, option_env!, panic!, print!, println!, select!, stringify!, thread_local!, try!, unimplemented!, unreachable!, vec!, write!, writeln!}  
, morekeywords=[6]{windows, enumerate, filter_map, iter, take_while, for_each}
, keywordstyle=\color{cobalt}
, keywordstyle=[2]\color[rgb]{0.75, 0, 0}
, keywordstyle=[3]\color[rgb]{0, 0.5, 0}
, keywordstyle=[4]\color[rgb]{0, 0.5, 0}
, keywordstyle=[5]\color[rgb]{0, 0, 0.75}
, keywordstyle=[6]\color{blue}
, stringstyle=\color{copper}%
, basicstyle=\tiny\ttfamily%
, numbers=left%
, numbersep=6pt%
, numberstyle=\tiny\ttfamily\color{gray}%
, breaklines=true%
, escapeinside={(*@}{@*)}%
, showstringspaces=false%
, xleftmargin=8pt%
}%

\lstdefinelanguage{myC}[]{C}{
  morekeywords={assert},
  morecomment=[f][\color{blue}]{@@},     
  morecomment=[f][\color{red}]-,         
  morecomment=[f][\color{ForestGreen}]+, 
  morecomment=[f][\color{magenta}]{---}, 
  morecomment=[f][\color{magenta}]{+++},
  morecomment=[f][\color{magenta}]{//},
  morecomment=[f][\color{magenta}]{/*},
  morecomment=[f][\color{magenta}]{\ \ \ \ //},
  morecomment=[f][\color{magenta}]{//},
  keywords = {DROP_POLICY_DENY, DROP_POLICY_AUTH_REQUIRED, CTX_ACT_OK,
    DROP_INVALID, BMC_MAX_PACKET_LENGTH, ENABLE_IPV4_FRAGMENTS,
    DROP_CT_INVALID_HDR, ENABLE_SRV6_SRH_ENCAP, IPPROTO_IPIP, IPPROTO_IPV6,
    ETH_HLEN, POLICY_MATCH_L4_ONLY, POLICY_MATCH_L3_PROTO, IS_ERR, likely,
    __always_inline, offsetof},
  keywords = [2]{long, int, u32, const, void, if, return, switch, return, case,
    break, u64, u8, unsigned, struct, sizeof, __u8, __s8, default, else, for,
    static, asm, volatile},
  keywordstyle=\color{purple},
  keywordstyle=[2]\color{blue},
  commentstyle=\color{darkspringgreen},
  stringstyle=\color{blue},
  numbers=left,
  basicstyle=\scriptsize\ttfamily,
  numbersep=6pt,
  numberstyle=\tiny\ttfamily\color{gray},
  breaklines=true,
  escapeinside={(*@}{@*)},
  showstringspaces=false,
  xleftmargin=8pt,
}

\lstdefinelanguage{myBPF}[]{C}{
  morecomment=[f][\color{magenta}];,
  keywords = {r1, r2, r3, r4, r5, r6, r7, r8, r9, ctx, pkt,
    w1, w2, w3, w4, w5, w6, w7, w8, w9},
  keywords = [2]{u8, u16, u32, u64},
  keywordstyle=\color{purple},
  keywordstyle=[2]\color{blue},
  commentstyle=\color{darkspringgreen},
  stringstyle=\color{blue},
  numbers=left,
  basicstyle=\scriptsize\ttfamily,
  numbersep=6pt,
  numberstyle=\tiny\ttfamily\color{gray},
  breaklines=true,
  escapeinside={(*@}{@*)},
  showstringspaces=false,
  xleftmargin=8pt,
}

\newenvironment{packed_itemize}{
    \begin{list}{\labelitemi}{\leftmargin=1.0em}
     \setlength{\itemsep}{2.5pt}
     \setlength{\parskip}{0pt}
     \setlength{\parsep}{0pt}
     \setlength{\headsep}{0pt}
     \setlength{\topskip}{0pt}
     \setlength{\topmargin}{0pt}
     \setlength{\topsep}{0pt}
     \setlength{\partopsep}{0pt}
    }{\end{list}}

%% file: abstract.tex
\begin{abstract}
\noindent
Safe kernel extensions
  have gained significant traction, 
  evolving from simple packet filters to
  large, complex programs that customize storage, networking,
  and scheduling.
Existing kernel extension mechanisms like eBPF
  rely on in-kernel verifiers to ensure safety of kernel extensions
  by static verification using symbolic execution.
We identify significant usability issues---safe extensions being rejected
  by the verifier---due to the
  {\em \gap{}}, a mismatch between developers' expectation of program
  safety provided by a contract with the programming language,
    and the verifier's expectation.

We present Rex, a new kernel extension framework that closes the \gap{}
  and improves the usability of kernel extensions in terms of
  programming experience and maintainability.
Rex builds upon language-based safety to provide safety
  properties desired by kernel extensions, along with a lightweight extralingual
  runtime for properties that are unsuitable for static analysis, including
  safe exception handling, stack safety, and termination.
With Rex, kernel extensions are written in {\it safe} Rust
  and interact with the kernel via a safe interface provided by Rex's kernel crate.
No separate static verification is needed.
\projname{} addresses usability issues of eBPF kernel
  extensions without compromising performance.
\end{abstract}

%% file: intro.tex
\section{Introduction}
\vspace{-5pt}


Kernel extensibility is an essential capability of modern
    Operating Systems (OSes).
Kernel extensions allow users with diverse needs to customize the OS without
    adding complexity to core kernel code or introducing disruptive
    kernel reboots.

In Linux, kernel extensibility has traditionally taken the form of loadable kernel
    modules. However, kernel modules are inherently {\it unsafe}---simple programming
        errors can crash the kernel.
Despite the support of safe languages like Rust~\cite{rust-for-linux-doc},
    there is no systematic support to ensure the safety of kernel modules---\textit{unsafe} Rust code 
    is allowed in kernel modules 
    wherein checks to prevent errors are non-existent.
Moreover, the vast, arbitrary interface exposed to kernel modules creates
    significant challenges in providing a safe Rust kernel
    abstraction to enforce safe Rust code~\cite{Miller-hotos19,rust-module-dev-quit-lwn}.


Recently, eBPF extensions have gained significant traction and
    become the {\it de facto}
    kernel extensions~\cite{cilium-docs,ebpf-windows}.
Core to eBPF's value proposition is a promise of {\it safety} of kernel extensions,
    enforced by the in-kernel verifier.
The verifier
    statically analyzes extension programs in eBPF bytecode,
    compiled from high-level languages (C and Rust).
It performs symbolic execution against every possible code path in the bytecode
    to check safety properties (e.g.,
    memory safety, type safety, termination, etc).
The kernel rejects any extension the verifier fails to verify.
Today, eBPF extensions have evolved far beyond simple packet filters (its original use cases~\cite{pf,bsdpf})
    and are increasingly being
    used to construct large, complex programs that customize storage~\cite{BMC,Zhong:osdi:2022,fetchbpf,lambda-io},
    networking~\cite{Electrode,DINT},
    CPU scheduling~\cite{ghost-scheduler,ghost-scheduler-lpc,sched-ext}, etc. 

However, we observe that eBPF's static verifier introduces
    significant usability issues, 
    making eBPF extensions
    hard to develop and maintain, especially for large, complex programs.
For example, the eBPF verifier often incorrectly
    rejects {\it safe} extension code due to fundamental limitations of static verification
    and defects in the verifier implementation.
When such false rejections happen,
    developers have no choice but to refactor or rewrite extension programs
    in ways that ``please'' the verifier.
Such efforts range from breaking an extension program into multiple small ones,
    nudging compilers to generate verifier-friendly code, tweaking code
    to assist verification, etc (see \S\ref{sec:motivation}).
Some of the efforts also involve reinventing wheels and hacking eBPF bytecode,
    which creates significant cognitive overheads and
    makes maintenance difficult.



We argue that these usability issues are rooted in the gap
    between the programming language and the eBPF verifier, which we term
    the {\em \gap{}}.
When writing eBPF programs, developers
    interact with the high-level language and naturally obey a
    {\em language contract} to align with the safety requirements of the language.
The compiler also adheres to the language contract.
Unfortunately, the verifier is not part of the language contract and
has different expectations.  As a result, verifier rejections may be
surprising; the feedback (verifier log) is at the bytecode level and is
    hard to map to source code.
As a result,
    developing eBPF extensions requires not only a deep knowledge
    of the high-level language and safety properties of kernel extensions
but also a deep understanding of implementation details and quirks of
    the verifier.

Unfortunately, recent efforts to improve the eBPF verifier (e.g., via testing and verification~\cite{formal-verifier-ebpf,lpc-24-agni})
    cannot fundamentally
    close the \gap{} because
    (1) they do not address scalability issues of symbolic execution, so
        extension programs have to ill-fit the verifier's internal limits,
    and
    (2) it is unlikely that the language compiler (e.g., LLVM) and the eBPF verifier
        are always in synchronization, given their independent developments.
Recent efforts to improve extension expressiveness via techniques
    like software fault isolation (e.g., KFlex~\cite{dwivedi-sosp24})
largely inherit the eBPF verifier and, therefore, do not address the \gap{}.

We present Rex, a new kernel extension framework that
    closes the \gap{} and effectively improves the usability of kernel extensions,
    in terms of programming experience and maintainability.
Rex builds up safety guarantees for kernel extensions based on safe language features.
With Rex, safety properties are checked by the language compiler within the language contract.
Rex drops the need for an extra verification layer and closes
    the \gap{}.
We choose Rust as the safe language, as it is already supported by Linux~\cite{rust-for-linux-doc}
    and offers desired language-based safety for practical systems programming~\cite{Miller-hotos19,rustSystems,theseus,redleaf}.


%
%




Rex kernel extensions are strictly written in {\it safe} Rust with selected
    features (unsafe Rust
    code is forbidden in Rex extensions).
Rex transforms the promises of Rust into safety guarantees for
    extension programs with the following endeavors.
First, to enable Rex extensions to be written entirely in safe Rust
    in the context of kernel extension,
    Rex develops a kernel crate and offers a safe kernel interface that
    wraps the existing eBPF kernel interface (eBPF
    helper functions and data types) with safe Rust wrappers and bindings.
The kernel crate enforces memory safety, extends type safety,
    and ensures safe interactions with the kernel.
\projname{} further enforces only safe Rust features through its compiler toolchains.


Moreover, \projname{} employs a lightweight extralingual runtime for
    safety properties that are hard to guarantee by static analysis.
Specifically, \projname{} supports safe stack unwinding and resource cleanup
    upon Rust panics at runtime.
\projname{} also checks kernel stack usage and uses
    watchdog timers to ensure termination with a safe mechanism.
The Rex runtime is engineered with minimal overhead to achieve
    high performance. 

We evaluate \projname{} on both its usability and performance.
We show that by closing the \gap{} and offering Rust's rich
    built-in functionality, Rex effectively rules out the usability
    issues in eBPF.
We further evaluate the usability by implementing the BPF Memcached Cache (BMC)~\cite{BMC}
    (a complex, performance-critical program written in eBPF)
    using Rex and show that Rex leads to cleaner, simpler extension code.
We also conduct extensive macro and micro benchmarks.
Rex extensions deliver the same level of performance as eBPF extensions---the
    enhanced usability does not come with a performance penalty.



\vspace{-2.5pt}
\para{Limitations.} 
\projname{}'s design comes with tradeoffs.
To close the \gap{} and improve usability, Rex requires kernel extensions to be written in Rust,
    though its design principles apply to other safe languages.
Rex brings the Rust toolchain into the Trusted Computing Base (TCB)
    and adds additional runtime complexity.
Note that Rex extensions and eBPF extensions can co-exist---Rex and eBPF
    represent different tradeoffs.
Rex targets large, complex kernel extensions for which usability
    and maintainability are critical.

\para{Contributions.} We make the following main contributions:
\begin{packed_itemize}
    \item A discussion of the \gap{}
      and its impact on the usability and maintainability of safe kernel extensions;
    \item Design and implementation of the \projname{} kernel extension framework,
        which closes the \gap{} by realizing safe kernel extensions
        upon language-based safety,
        together with efficient runtime techniques;
    \item The Rex project is at \url{https://github.com/rex-rs}.
\end{packed_itemize}

%% file: background.tex
\section{Safety of Kernel Extensions}
\label{sec:background}
\vspace{-5pt}

Safety is critical to OS kernel extensions---extension 
  code runs directly in kernel space, and bugs 
  can directly crash a running kernel. 
%
The eBPF verifier checks safety properties of  
  extension programs in bytecode 
  before loading them into the kernel to prevent
  programming errors such as illegal memory access. 
The verifier also checks the extension's interactions with the kernel via a 
  bounded interface, defined by eBPF {\em helper functions},
  to prevent resource leaks and deadlocks. 
We summarize the safety properties targeted by eBPF as follows:
\begin{packed_itemize}
  \vspace{-5pt}
\item {\bf Memory safety.} Kernel extensions can only access pre-allocated memory
  via explicit context arguments or kernel
  interface (helper functions), preventing NULL pointer dereferencing
  and corruption of kernel data structures.
\item {\bf Type safety.} When accessing data in memory, kernel extensions must use
  the correct types of data, avoiding misinterpretation of the data
  and memory corruption.
\item {\bf Resource safety.} When acquiring kernel resources
  (e.g., locks, memory objects, etc.) through helper functions,
  kernel extensions must eventually invoke the appropriate interface
  to release the resources, preventing memory leaks or
  deadlocks that can crash or hang the kernel.
\item {\bf Runtime safety.} Kernel extensions must terminate, with no
  infinite loops that can hang the kernel indefinitely.
\item {\bf No undefined behavior.} Kernel extensions must never
  exhibit undefined behavior, such as integer errors (e.g.,
  divide by zero) that cause kernel crashes. 
\item {\bf Stack safety.} 
  Kernel extensions must not overflow the limited and fixed-size kernel stack,
  avoiding kernel crashes or kernel memory corruptions.\vspace{-5pt}
\end{packed_itemize}

Note that the above notion of safety in eBPF focuses on preventing 
  programming errors that may crash or hang the kernel.
Despite the discussions on whether security is a reasonable 
  target~\cite{beebox-security24,safebpf-thomas},
  in practice, eBPF and other extension frameworks (e.g., KFlex~\cite{dwivedi-sosp24}) no longer pursue
  unprivileged use cases due to its inherent limitations (see detailed discussion in \S\ref{sec:safety_model})~\cite{reconsider-unpriv-ebpf-lwn,ebpf-sec-lwn,pawan-8a03e56b253e}. 
Our work follows this safety model.

%

%% file: motivation.tex
\vspace{-7pt}
\section{The Language-Verifier Gap}
\label{sec:motivation}
\vspace{-5pt}

\begin{figure}
    \includegraphics[width=1.0\linewidth]{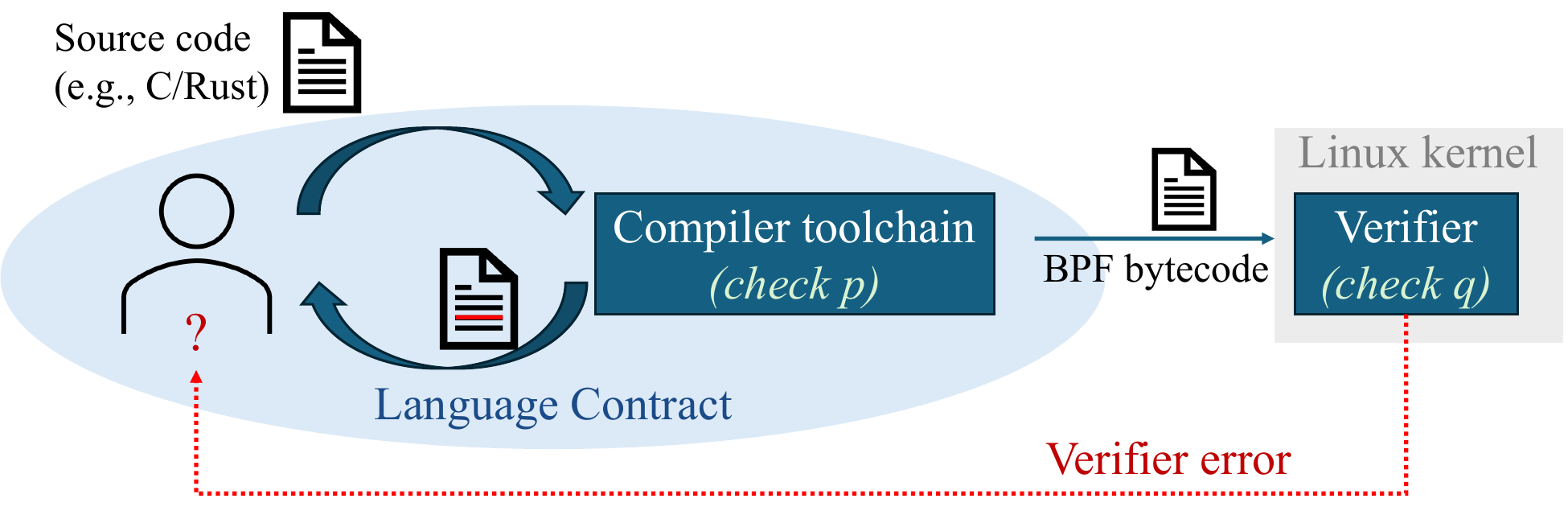}
    \centering
    \vspace{-15pt}
    \caption{The \gap{}}
    \label{fig:gap}
\end{figure}

A fundamental problem of eBPF's safety verification mechanism is the {\it \gap{}}
    (illustrated in Figure~\ref{fig:gap}).
Developers mainly implement and maintain eBPF extensions in high-level
    languages (e.g., C and Rust) and compile them to eBPF bytecode;
they agree to a contract with the high-level language (code has property $p$), which is
    enforced by the compiler.
When a program fails to compile, developers receive feedback
    about how they violate
    the language contract.
However, in eBPF, the compiled extension code will be further checked by the verifier.
If correctly compiling code fails
    to pass the verifier (e.g., the code lacks property $q$, which is not
    part of the contract),
    it is difficult for developers
    to understand why the extension program fails despite obeying the contract.


The \gap{} is further exacerbated when the verifier incorrectly rejects {\it safe} extension programs
    due to (1) scalability limitations of the symbolic execution used by the verifier,
    (2) conflicting analyses between the compiler and the verifier,
    and (3) the verifier's implementation defects.
Today, the \gap{} forces developers to understand the verifier's internal implementation and
    its limitations and defects,
    and to revise extension code in ways that can pass the verifier at the bytecode level.
Many such revisions are workarounds solely to
    please the verifier.
Fundamentally, the \gap{} breaks the language abstractions
    and artificially forces extension programs
    in a high-level language
    to tightly couple with the low-level verifier implementation.

\vspace{-2pt}
\subsection{Verifier Workarounds}
\label{sec:workaround}
\vspace{-2pt}

To understand the impact of the \gap{}, 
we analyzed commits related to revising eBPF programs to resolve verifier issues
    in popular eBPF projects, including Cilium~\cite{cilium}, Aya~\cite{aya-rs}, and
Katran~\cite{katran}.
The commits were collected by searching through the commit logs of each project
    using keywords and manually inspected. 
In total, we collected 
72 commits related to verifier issues.
We also included two issues raised by BMC~\cite{BMC} and Electrode~\cite{Electrode}
    in the discussion. 

In all 72 commits, we confirmed that the original eBPF programs were safe
    but were rejected by
    the verifier due to defective or overly conservative safety checks.
When the verifier rejects a safe program, the developer must
find a {\em workaround}.
Table~\ref{fig:commit-table} summarizes the workaround patterns.
\para{Refactoring extension programs into small ones.}
\label{motivation:restructure}
The most common pattern (27 out of 72) is refactoring a large eBPF program, which
    the verifier rejects due to exceeding the verifier's internal limits
    into smaller ones.
Since symbolic execution is hard to scale,
    the eBPF verifier imposes a series of limits on the complexity of extension programs
    (e.g., the number of bytecode instructions and branches~\cite{starovoitov-51580e798cb6})
    to ensure
    verification to complete at load time.
The eBPF extension will be rejected if it exceeds any of these limits.
Such rejections have no implication on the safety of the extension;
rather, they are artifacts of scalability limitations
    of static verification.

We observe two standard practices of refactoring eBPF programs to work around verifier limits:
    (1) splitting eBPF programs into smaller ones
    and (2) rewriting eBPF programs with reduced complexity the verifier can handle.

We use BMC~\cite{BMC} as an example to explain these practices.
BMC uses eBPF to implement in-kernel caches to accelerate Memcached.
Conceptually,
    only two extension programs are needed (at ingress and egress, respectively).
However, to satisfy the verifier limit, BMC developers had to split BMC code into {\it seven}
    eBPF programs connected via tail calls.\footnote{Since BMC, the limit has
        increased, but the fundamental gap remains.}
Such splitting creates an unnecessary burden on the implementation and maintenance of BMC;
    it also creates performance issues when states need to pass across tail calls (using maps).

\begin{table}[t]
    \small
    \caption{Patterns of common verifier workarounds} 
    \vspace{-5pt}
    \label{fig:commit-table}
    \centering
    \begin{tabular}{lc}
        \toprule
        \textbf{Category} & \textbf{Count} \\
        \midrule
        Refactoring extension programs into small ones & 27 \\           
        Hinting LLVM to generate verifier-friendly code & 22 \\          
        Changing code to assist verification & 15 \\              
        Dealing with verifier bugs & 9 \\          
        Reinventing the wheels & 1 \\              
        \bottomrule
    \end{tabular}
    \vspace{5pt}
\end{table}



Despite the smaller size of each program after splitting, BMC programs that
    iterate over the packet payload in a loop cannot easily pass the verifier.
While the programs correctly check for the bounds of the payload,
    the programs result in an excessive number of jump instructions
    and exceed the verifier's complexity limit.
As a workaround, developers must bind the size of the data BMC can
    handle further to pass the verifier.
\S\ref{eval:bmc-case-study} revisits this example in more depth.

\para{Hinting LLVM to generate verifier-friendly code}
\label{motivation:llvm-codegen}
Another common pattern is to change source code in ways that
    nudge the compiler (LLVM) to generate verifier-friendly bytecode.
In several cases, LLVM generated eBPF bytecode that fails the verifier
    due to complex, often undocumented expectations of the verifier.
Figure~\ref{fig:inline-error} shows a case from Cilium~\cite{chaignon-847014aa62f9}
    that accesses a pointer field
    (\texttt{\small ctx->data}) in a socket buffer, defined as a 32-bit
    integer in the kernel \texttt{\small uapi} interface.
\new{
LLVM generates a 32-bit load on \texttt{\small data} and assigns its value to
    another 32-bit register.
While \texttt{\small data} is defined as 32-bit, under the hood it represents a
    pointer to the start of the packet payload.
}
The 32-bit assignment made the verifier interpret the pointer
    as a scalar and incorrectly reject the program when it tries to access
   memory through the scalar.
As a workaround, developers encapsulated access to
    \texttt{\small data} in inline assembly
    (Figure~\ref{fig:inline-asm}) to prevent LLVM from generating
    32-bit move as an optimization (LLVM does not optimize
    inlined assembly).
The verifier then treats the register as a pointer rather than a
    scalar.






\begin{figure}[tp]
\begin{subfigure}{0.48\textwidth}
\begin{lstlisting}[language=myBPF]
; return (void *)(unsigned long)ctx->data;
2: (61) r9 = *(u32 *)(r7 +76)
; R7_w=ctx(id=0,off=0,imm=0) R9_w=pkt(id=0,off=0,r=0,imm=0)
3: (bc) w6 = w9
; R6_w=inv(id=0,umax_value=4294967295,var_off=(0x0; 0xffffffff))
; R9_w=pkt(id=0,off=0,r=0,imm=0)
...
7: r2 = *(u8 *)(r6 +22)
; R6 invalid mem access 'inv'
\end{lstlisting}
    \vspace{-5pt}
    \caption{Verifier log showing an invalid memory access, which is hard to
        diagnose and does not directly map to the source code in C}
    \vspace{-10pt}
    \label{fig:inline-error}
\end{subfigure}
\begin{subfigure}{0.48\textwidth}
\begin{lstlisting}[language=myC]
static __always_inline void *
ctx_data (const struct __sk_buff *ctx) {
  void *ptr;
  asm volatile (
    "%0 = *(u32 *)(%1 + %2)"
    : "=r"(ptr)
    : "r"(ctx),
      "i"(offsetof(struct __sk_buff, data))
  );
  return ptr;
}
\end{lstlisting}
    \vspace{-5pt}
    \caption{Inline assembly code created to work around the verification failure
        by preventing the compiler optimization}
    \vspace{-5pt}
    \label{fig:inline-asm}
\end{subfigure}
\caption{An example of the \gap{} from Cilium~\cite{chaignon-847014aa62f9}, where a safe eBPF extension is incorrectly rejected by the verifier
    (\ref{fig:inline-error})
    and developers had to work around the problem by creating inline assembly code (\ref{fig:inline-asm}).}
\label{fig:llvm-cg-issue}
\end{figure}

In another case~\cite{borkmann-394e72478a8d},
    developers were forced to use \texttt{\small volatile} when
    loading from a 32-bit integer pointer and only using its upper 16 bits.
Without \texttt{\small volatile}, LLVM optimized the code to only load the
    upper 16-bit from the pointer, which the verifier perceives as a size
    mismatch violation.

In fact, many eBPF programs today can only pass the verifier if compiled with
    \texttt{\small -O2} optimization---the verifier has a hardwired
    view of eBPF extension bytecode, which the compiler cannot
    generate with other levels, including \texttt{\small -O0}.


\para{Changing code to assist verification.}
\label{motivation:add-code}
In this pattern, developers had to
    assist the verifier manually. 
A common pattern is refactoring the code into new functions when
    the verifier loses track of values in eBPF programs.
It is often unclear what code needs to be refactored to pass the
    verifier, which significantly burdens developers.
Figure~\ref{fig:inline-fig} shows a code example from
    Cilium~\cite{rajahalme-847014aa62f9}, which originally
    used a \texttt{\small goto} statement to combine the code path of
    \texttt{\small policy} and \texttt{\small l4policy} to avoid duplicated code.
However, the combined code, which assigns \texttt{\small l4policy} to
    \texttt{\small policy}, later causes the verifier to incorrectly believe
    that \texttt{\small policy}, which is a pointer variable, is instead a
    scalar and reject the program.
As a workaround, developers had to refactor the policy check code
    into an inlined function to separate the code path to pass the verifier.

\begin{figure}[tp]
\begin{subfigure}{0.48\textwidth}
\begin{lstlisting}[language=myC]
  if (likely(l4policy && !l4policy->wildcard_dport)) {
    *match_type = POLICY_MATCH_L4_ONLY;
-   policy = l4policy;
-   goto policy_check_entry;
+   return __account_and_check(ctx, l4policy, ...);
  }

  if (likely(policy && !policy->wildcard_protocol)) {
    *match_type = POLICY_MATCH_L3_PROTO;
-   goto policy_check_entry;
+   return __account_and_check(ctx, policy, ...);
  }

\end{lstlisting}
    \vspace{-10pt}
    \caption{Simplifying control flow}
    \vspace{-10pt}
    \label{fig:inline-fig}
\end{subfigure}
\begin{subfigure}{0.48\textwidth}
\begin{lstlisting}[language=myC]
  #ifndef ENABLE_SKIP_FIB
  ...
  if (likely(ret == BPF_FIB_LKUP_RET_NO_NEIGH)) {
    nh_params.nh_family = fib_params->l.family;
    ...
  } else {
    return DROP_NO_FIB;
  } ...
  skip_oif:
  #else
  *oif = DIRECT_ROUTING_DEV_IFINDEX;
- nh_params.nh_family = fib_params->l.family;
  #endif /* ENABLE_SKIP_FIB */
  ...
- dmac = nh_params.nh_family == AF_INET ? ...;
+ dmac = fib_params->l.family == AF_INET ? ...;
\end{lstlisting}
    \vspace{-5pt}
    \caption{Simplifying data flow}
    \vspace{-5pt}
    \label{fig:teach-verifier}
\end{subfigure}
\caption{Examples that developers had to work around the \gap{} by refactoring
    already safe extensions}
    \label{fig:inline-fig}
\end{figure}

Developers also have to teach the verifier by
    providing additional information.
Figure~\ref{fig:teach-verifier} shows an example in Cilium~\cite{borkmann-efb5d6509fea}
    where the verifier lost track of \texttt{\small nh\_params.nh\_family},
    a scalar spilled onto the stack and mistakenly treated it as a
    pointer when loading it back, leading
    to an invalid size error on the load.
As a workaround, developers passed \texttt{\small fib\_params->l.family} directly instead of
    going through \texttt{\small nh\_params.nh\_family} to let the verifier
    know the scalar value. 



\para{Dealing with verifier bugs.}
\label{motivation:kernel-version}
The \gap{} is further exacerbated by verifier bugs~\cite{untenableVerification,formal-verifier-ebpf,proof-carrying-verifier,sandbpf},
    as developers need to
    acquire knowledge of the verifier's expectations and deficiencies.
Moreover, different kernel versions can have different verifier bugs.
Dealing with verifier bugs and maintaining
    compatibility across kernel versions is non-trivial.
In a Cilium case~\cite{graf-e38a92115620},
The verifier rejected a correct program with valid access to the context
    pointer due to the verifier's incorrect handling of constant pointer offsets.
The verifier bug was known, but the fix was not present in
    all kernel versions.
Cilium developers had to tweak their program to avoid the
    bug-triggering yet correct context pointer access so the code could verify on
    all kernel versions.

\para{Reinventing the wheels.}
\label{motivation:wheel}
Developers may need to reimplement existing functions
    to pass the verifier.
In Aya, the default definition of the
    \texttt{\small memset} and \texttt{\small memcpy} intrinsics provided by
    the language toolchain failed
    to pass the verifier~\cite{aya-pr-698}.
Aya eventually implemented its own version for both intrinsics, using a simple
    loop to iterate over the data to avoid ever tripping the verifier.
This case reflects a key challenge of using eBPF for large, complex extension programs,
    as developers may need to re-implement many standard, nontrivial library functions.


\vspace{-5pt}
\subsection{Implications}
\vspace{-5pt}

Our analysis shows that the \gap{} causes severe usability issues
    in developing and maintaining
    eBPF kernel extensions. 
eBPF developers have to implement arcane fixes and change their mental
    model to meet the verifier's constraints. 
If an eBPF extension fails to verify, the verifier log rarely pinpoints the
    root causes and cannot help trace back to the source code.
Since it is hard to require compilers like LLVM
    to follow the eBPF verifier's implementations,
    we expect the \gap{} will continue to exist, 
    especially for large, complex extension programs.

%% file: safety_model.tex
\section{Key Idea and Safety Model}
\label{sec:safety_model}
\vspace{-5pt}

The key idea of Rex is to realize {\it safe} kernel extensions without a separate layer of
    static verification.
Our insight is that the desired safety properties of kernel extensions
    can be built on the foundation of language-based properties of
    a safe programming language like Rust,
    together with extralingual runtime checks.
In this way, the in-kernel verifier can be dropped, and
    the \gap{} can be closed.
Rex extensions are strictly written in a {\it safe} subset of Rust.
We choose Rust as the safe language for kernel extensions
    (instead of other languages like Modula-3~\cite{spin}
    and Sing\#~\cite{lang-sing})
    because Rust is already supported
    by Linux~\cite{rust-for-linux-lwn} and offers
    desired language features for practical kernel code~\cite{redleaf,theseus,tockos}.
\new{Rex enforces the same set of safety properties eBPF enforces (\S\ref{sec:background}).}
Hence, Rex extensions fundamentally differ from unsafe kernel modules.

\para{Safety Model.}
Rex follows eBPF's non-adversarial safety model---the
    safety properties focus
    on preventing programming errors from crashing/hanging the kernel instead of
    malicious attacks.
Like eBPF, Rex extensions are installed from a trusted context with root
    privileges on the system.
Rex extensions can only be written in safe Rust with selected features
    and language-based safety is enforced out by a trusted Rust compiler (\S\ref{sec:lang_subset}). 
Unlike Rust kernel modules that can use 
    unsafe Rust, the language-based safety of Rex extensions is strictly enforced.
Other safety properties that are not covered by language-based safety (e.g., termination)
    are checked and enforced by the lightweight Rex runtime.

While historically eBPF supported unprivileged mode~\cite{reconsider-unpriv-ebpf-lwn} and
    there are research efforts in supporting
    unprivileged use cases for kernel extensions~\cite{beebox-security24,safebpf-thomas,jia2023},
in practice, eBPF and other frameworks (e.g., KFlex~\cite{dwivedi-sosp24}) no longer pursue
    it~\cite{ebpf-sec-lwn,pawan-8a03e56b253e}.
The reasons come from inherent limitations of securing eBPF
    or kernel extensions in general.

First, it is hard for the eBPF verifier to prevent transient execution attacks like Spectre attacks
    completely, without major performance and compatibility overheads (see~\cite{ebpf-sec-lwn}).
Specifically, new Spectre variants are being discovered; though many of them are bugs in hardware,
    they cannot be easily detected and fixed by static analysis~\cite{perspective_isca}.
Sandboxing techinques cannot completely prevent Spectre attacks either,
    e.g., SafeBPF~\cite{safebpf-thomas} only prevents memory vulnerabilities,
    while BeeBox~\cite{beebox-security24} only focuses on two Spectre variants and requires manual instrumentation of helper functions.
For these reasons, the Linux kernel and major distributions also have moved away from unprivileged
    eBPF~\cite{pawan-8a03e56b253e,unpriv-ebpf-ubuntu,unpriv-ebpf-suse}.

Second, eBPF chose not to be a sandbox environment (like WebAssembly or JavaScript)
    that does not know what code will be run~\cite{ebpf-sec-lwn}.
Instead, the development of eBPF assumes that ``{\it the intent of a BPF program is known}~\cite{ebpf-sec-lwn}.''

Lastly, the constantly reported verifier vulnerabilities~\cite{untenableVerification,ebpf-stackoverflow,ebpf-termination}
    indicate that a bug-free verifier is hard in practice.





%

\para{Trusted Computing Base (TCB).}
With Rex's safety model, the TCB consists of the Rust toolchain, the
    \projname{} kernel crate, and the \projname{} runtime.
\projname{} has to trust the Rust toolchain for its correctness to deliver
    language-based safety.
We believe the need to trust the Rust toolchain is acceptable
    and does not come with high risks with our safety model.
Recent work on safe OS kernels~\cite{theseus,redleaf,Miller-hotos19} makes the same decision
    to establish language-based safety by trusting the Rust toolchains.
The active effort on extensive fuzzing and formal verification of the Rust
    compiler~\cite{rust-belt,stacked-borrows-popl19,verus,verus-sosp24,rvt,rustc-fuzzing}
    may further reduce the risk.
Certainly, we acknowledge that the existing Rust compiler, such as rustc~\cite{rustc},
    is larger than the eBPF verifier.

%% file: principle.tex
\vspace{-8pt}
\section{\projname{} Design}
\label{sec:principle}
\vspace{-5pt}

\begin{figure}
    \centering
    \includegraphics[width=1.0\linewidth]{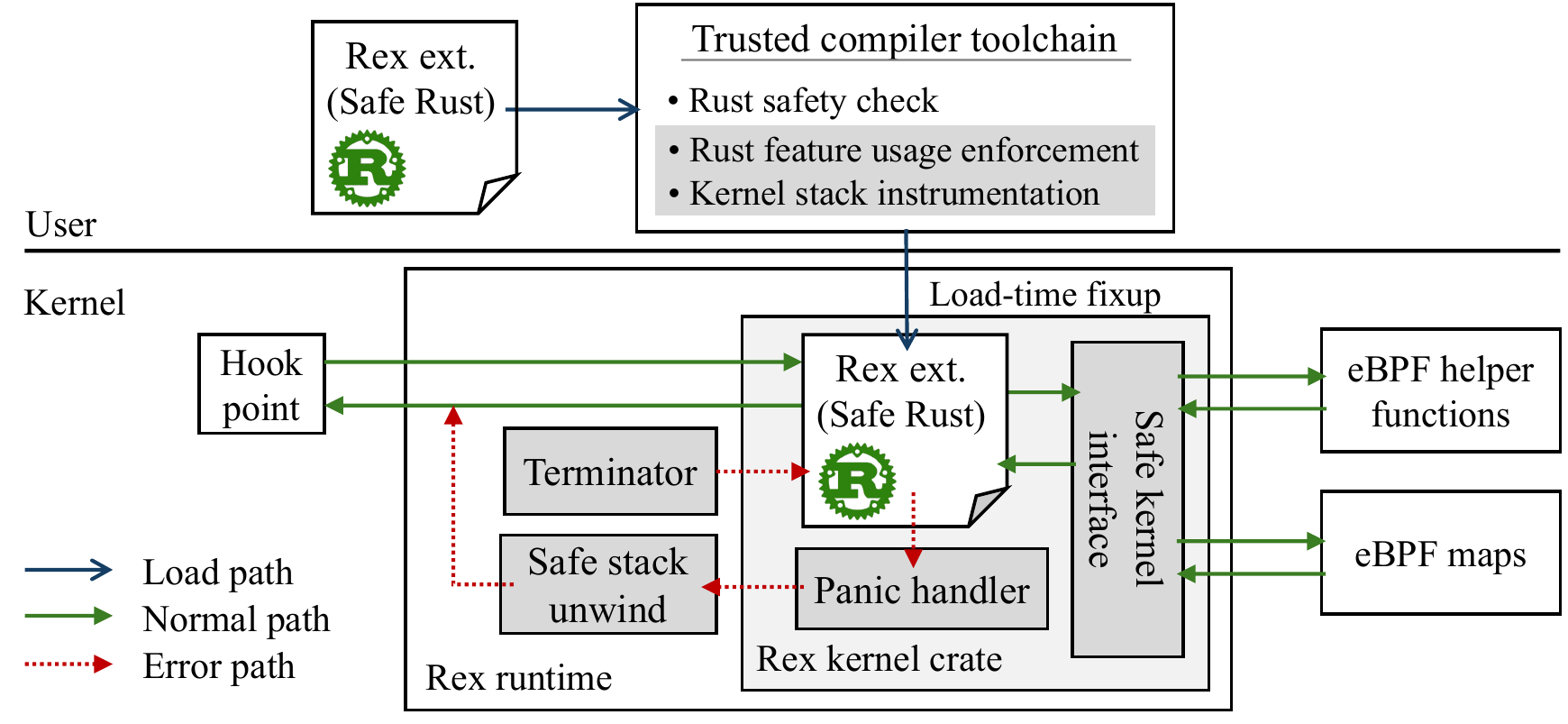}
    \vspace{-15pt}
    \caption{Overview of the \projname{} kernel extension framework.
    The gray boxes are Rex components.}
    \label{fig:rex-overview}
    \vspace{-5pt}
\end{figure}

The key challenge of the Rex design is to provide safety guarantees of kernel extensions
    (listed in \S\ref{sec:background})
    on top of Rust's safe language features (adopting a safe language alone is insufficient,
    as in Rust kernel modules).

Figure~\ref{fig:rex-overview} gives an overview of the \projname{} framework.
To realize language-based safety, Rex enforces kernel extensions to be strictly
    written in {\it safe} Rust with selected features.
The \projname{} compiler toolchain rejects any Rex program that uses unsafe
    language features.
\new{Although this safe subset of Rust already provides inherent language-based
    safety within Rex extensions, eliminating undefined behaviors,
    safety of extensions is only achieved with the presence of safe kernel
    interactions provided by the \emph{\projname{} kernel crate}.
The kernel crate is trusted and
    bridges Rex extensions with unsafe kernel code.}
Rex builds on top of the eBPF helper functions interface to provide a safe
    kernel interface for Rex extensions to interact with the kernel
    using safe Rust wrappers and bindings.
The safe interface encapsulates the interaction across the foreign
    function interface in the kernel crate.
\new{We reuse the eBPF helper interface, because it is designed
    for kernel extensions with a clearly defined programming contract and
    separates extensions from kernel's internal housekeeping
    (e.g., RCU~\cite{rcu}).}

Rex employs a lightweight extralingual in-kernel runtime that checks safety properties
    that are hard for static analysis. 
\new{The runtime enforces program termination, kernel stack safety, and safe
    handling of runtime exceptions (e.g., Rust panics).}

\vspace{-8pt}
\subsection{Safe Rust in \projname{}}
\label{sec:lang_subset}
\vspace{-3pt}

\projname{} only allows language features that are safe in the context of
    kernel extensions.
First, Rex excludes any \textit{unsafe} Rust code as it misses
    important safety checks
    from the Rust compiler and can voilate various safety properties (\S\ref{sec:background}).
Second, \projname{} also forbids Rust
    features that interfere with Rust's automatic management of object
    lifetimes, which include \texttt{\small core::mem::\{forget,ManuallyDrop\}}
    and the \texttt{\small forget} intrinsic.
\new{These features are considered safe in Rust but
    violate resource safety of kernel extensions by facilitating resource
    leakage.}
Third, language features that cannot be supported in the kernel extension context are
    excluded by \projname{}.
This group contains the \texttt{\small std}~\cite{rust-std} library and
    dynamic allocation support (not available in \texttt{\small no\_std} build~\cite{rust-nostd}), the
    floating point and SIMD support (generally cannot be used in kernel space),
    and the \texttt{\small abort} intrinsic (triggers an invalid instruction).
Note that dynamic allocation may be supported by hooking
    \texttt{\footnotesize alloc}~\cite{rust-alloc} crate to the kernel
    allocator.
\new{
We plan to explore the use of dynamic allocation in kernel extensions in
    future works (\S\ref{sec:discussion}).
}


To enforce the restrictions, \projname{} configures the Rust compiler and linter
    to reject the use of prohibited features.
Specifically, we set compiler flags~\cite{rustc-lint} to forbid unsafe code and
    unstable Rust features which includes SIMD and intrinsics.
For individual langauge items such as \texttt{\small core::mem::forget}, we
    configure the Rust linter~\cite{clippy-lints} to detect and deny their usage.
We further remove floating point support by setting the target
    features~\cite{rustc-codegen} in \projname{} compilation.
The \texttt{\small std} library and dynamic allocation are already unavailable
    in \texttt{\small no\_std} environment used by \projname{} and therefore
    warrant no further enforcement.

\vspace{-4pt}
\subsection{Memory safety}
\label{principle:memsafety}
\vspace{-3pt}

Rex enforces extensions to access kernel memory safely.
There are two common memory access patterns, depending
    on the ownership of the memory region:
    (1) memory owned by the extension (e.g., a stack buffer) is sent to the kernel
    through helper functions,
    and (2) memory owned by the kernel (e.g., a kernel struct) is accessed by
    the extension.

\para{Memory owned by extensions.}
A Rex extension can allocate memory on the stack
    and send it to the kernel (e.g., asking the kernel to fill a
    stack buffer with data) via existing eBPF helper functions.
Rex ensures no unsafe memory access and thus prevents stack buffer overflow and
    kernel crash (e.g., corruption of the return address on the stack).


Unlike eBPF that checks a memory region with its size of
    every invocation of helper functions,
in \projname{}, the strict type system of Rust already prevents unsafe access.
\projname{} leverages the generic programming feature of
    Rust~\cite{rust-generics} to ensure that the size sent through the helper
    function interface is always valid.
For helper functions that take in pointer and size as inputs,
    the \projname{} kernel crate creates an
    adaptor interface that parametrizes the pointer type as a generic type parameter.
The interface queries the size of the generic type from the compiler
    and invokes the kernel interface with this size as an argument.
Since Rust uses {\it monomorphization}~\cite{rustc-monomorphize}, the concrete
    type and its size are resolved at compile time, adding no runtime overhead.
In this way, the size is guaranteed to match the type statically and the
    kernel will never make an out-of-bound access.
This works for both scalar types and array types.
We use Rust's {\it const generics} to allow a constant to be used as a
    generic parameter~\cite{rust-generics} to encode array lengths.

\para{Memory owned by the kernel.} The kernel can provide
    extensions with a pointer to
    kernel memory (e.g., map value
    pointers and packet pointers).
The extension must not have out-of-bound memory access. 
In eBPF, the verifier checks uses of kernel pointers with a static size,
    e.g. map value pointers (maps store the size of values);
for pointers without a static size like packet data pointers, the verifier
    requires extensions to explicitly check memory boundaries.

In \projname{}, pointers with static sizes are handled through the Rust type
    system.
The kernel map interface of \projname{} encodes the key and value
    types through generics and returns such pointers to extension programs as
    safe Rust references.
To manage pointers referring to dynamically sized memory regions,
    the \projname{} kernel crate abstracts such pointers into a Rust
    \emph{slice} with dynamic size.
Rust slice provides runtime bounds checks (\S\ref{principle:eh}), which allows
    the check to happen 
    without explicit handling by the
    extension.


Rust slices are in principle similar to \texttt{\small dynptr}
    in eBPF~\cite{ebpf-dynptr-lwn}, but provide more flexibility.
eBPF \texttt{\small dynptr}s are pointers to dynamically sized data regions
    with metadata (size, type, etc); however,
    access to the \texttt{\small dynptr}'s referred memory must be of a static size.
Rust slices allow dynamically sized access to the underlying
    memory, benefiting from its runtime bounds checks.
Moreover, the \texttt{\small bpf\_dynptr\_\{read,write\}} helpers do not
    implement a zero-copy interface available in Rust slices.
While \texttt{\small bpf\_dynptr\_\{data,slice\}} helpers
    allow extensions to obtain data slices without copying, they
    again require explicit checks of the bound of the slice.
As a tradeoff, eBPF \texttt{\small dynptr}s avoids runtime overheads of
    dynamic bounds checks, which we find negligible in our evaluation (\S\ref{eval:macro}).

\vspace{-4pt}
\subsection{Extended type safety}
\vspace{-3pt}

Rex extends Rust's type safety to allow extension programs to safely convert a byte stream
    into typed data.
The pattern is notably found in networking use cases, where extensions need
    to extract the protocol header from a byte buffer in the packet as a struct.
Safety of such transformations is beyond Rust's native type safety
    because they inevitably involve unsafe type casting.
eBPF allows pointer casting;
the verifier ensures: (1) the program does not make a
    pointer from a scalar value, and (2) the new type fits the memory boundary.

\projname{} also enforces the above two properties so that
    the reinterpreting cast (dubbed ``transmute'' in Rust) is safe.
\projname{} extends Rust's type safety to cover such casts.
To do so,
\projname{} defines a set of primitive scalar types that are
    considered safe as targets for casting.
\projname{} requires the target type of casting to be of either one of the safe types
    or a structure type in which all the members are of the safe types.
The safe types are specified by implementing the \texttt{\small Rex::SafeTransmute}
    trait, which is sealed and only implementable from within the kernel crate.
We use the \emph{procedural macros}~\cite{rust-proc-macro} feature of Rust to enforce
    this constraint and generate the safe transmute interface at compile time on
    extended types.
Under the safe-transmute contraints, mismatched scalar types can only cause
    logical errors, but do not constitute safety violation.


\vspace{-4pt}
\subsection{Safe resource management}
\vspace{-3pt}

Rex extensions are ensured to acquire and release
    resources properly to avoid leaks of kernel resources (e.g., refcounts
    and spinlocks).
Different from eBPF where the verifier checks all possible code paths
    to ensure the release of acquired resources,
\projname{} uses Rust's Resource-Acquisition-Is-Initialization
    (RAII) pattern~\cite{rust-raii}---for every kernel resource
    a Rex extension may acquire, the \projname{} kernel crate defines an RAII wrapper type
    that ties the resource to the lifetime of the wrapper object.

For example, when the program obtains a spinlock from the kernel, the
    \projname{} kernel crate constructs and returns a \emph{lock guard}.
The lock guard implements the RAII semantics through the
    \texttt{\small Drop} trait~\cite{rust-drop} in
    Rust, which defines the operation to perform when the object is destroyed.
In the case of lock guard, its \texttt{\small drop} handler releases the lock.
\new{
\projname{} uses compiler-inserted \texttt{\small drop} calls at the end of
    object lifetime during normal execution, and implements its own resource
    cleanup mechanism (\S\ref{principle:eh}) for exception handling.
}
The use of RAII automatically manages kernel resources to ensure
    safe acquisition and release.
Extension programs do not need to explicitly release the lock or drop the lock
    guard.

\vspace{-4pt}
\subsection{Safe exception handling}
\label{principle:eh}
\vspace{-3pt}

While certain Rust safety properties are enforced statically by the compiler,
    the others are checked at runtime and their violations trigger exceptions (i.e., Rust panics).
To handle exceptions in userspace, Rust uses the Itanium exception handling ABI~\cite{itanium-abi} to
    unwind the stack.
A Rust panic transfers the control flow to the stack unwinding
    library (e.g., llvm-libunwind), which backtracks the call stack and executes
    cleanup code and catch clauses for each call frame.
Unfortunately, this ABI is unsuitable for kernel extensions:
\begin{packed_itemize}
    \item Unlike in userspace
        where failures during stack unwinding
        crash the program,\footnote{Theseus~\cite{theseus} implements stack unwinding in the kernel.
        But, it assumes that unwinding never fails; faults in unwinding result in kernel failures.}
    stack unwinding in kernel extensions cannot fail---kernel
        extensions must not crash the kernel and must not leak kernel resources.
    \item Unwinding generally executes destructors for all existing objects on
        the stack, but executing untrusted, user-defined destructors (via the
        \texttt{\small Drop} trait~\cite{rust-drop} in Rust) is unsafe.
\end{packed_itemize}
Rex implements its own exception handling framework with two main components: (1) graceful exit
    upon exceptions, which resets the context, and (2) resource cleanup to
    ensure release of kernel resources (e.g., reference counts and locks).

\begin{figure}
    \includegraphics[width=0.95\linewidth]{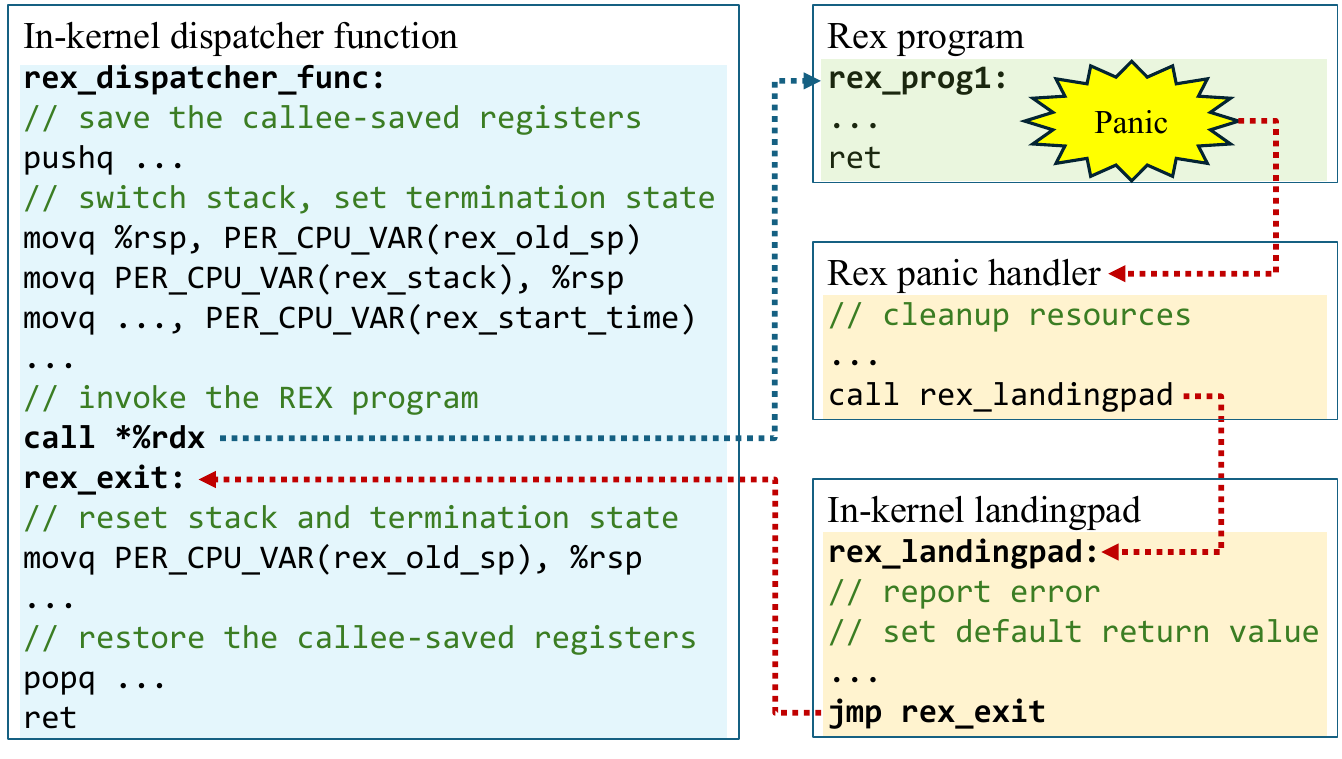}
    \centering
    \caption{Exception handling control flow in \projname{}}
    \label{fig:eh-overview}
    \vspace{-5pt}
\end{figure}

\para{Graceful exit.}
To ensure a graceful exit from an exception, \projname{} implements a small
    runtime (Figure~\ref{fig:eh-overview}) in the kernel, which
    consists of a program dispatcher, a panic handler, and a landingpad.
The dispatcher takes the duty of executing the extension program
    (like the eBPF dispatcher).
It saves the stack pointer of the current context into per-CPU
    memory, switches to the dedicated program stack (\S\ref{principle:stack}),
    sets the termination state (\S\ref{principle:termination}), and
    then calls into the program.
If the program exits normally, it
    returns to the dispatcher, which switches the stack back and clears the
    termination state.
Under exceptional cases where a Rust panic is triggered, the panic handler
    releases kernel resources currently allocated by the extension, and
    transfer control to the in-kernel landingpad to print
    debug information to the
    kernel ring buffer and return a default error code to the kernel.
Then, the landingpad redirects control flow to a pre-defined label
    in the middle of the dispatcher, where it restores the old value
    of the stack pointer from the per-CPU storage.
This effectively unwinds the stack and resets the context as if the extension returned successfully.

\para{Resource cleanup.}
Correct handling of Rust panics requires cleaning up resources acquired
    by the extension.
However, static approaches that rely on the verifier to pre-compute resources
    to be released during verification
    (e.g., object table in~\cite{dwivedi-sosp24}) do not apply to Rex due to the
    \gap{}.

Our insight is that extensions can only obtain
    resources by explicitly invoking helper functions. 
So, Rex records the allocated kernel resources
    during execution in a per-CPU buffer, which is in principle like the global
    heap registry in~\cite{redleaf}.
Upon a panic, the panic handler takes the responsibility to correctly
    release kernel resources, which involves traversing the
    buffer and dropping recorded resources.

\new{
We implement the cleanup code as part of the panic handler in the \projname{}
    kernel crate, as it
    is responsible for coordinating helper function calls
    that obtain kernel resources.
Implementing the cleanup mechanism in the kernel crate ensures safety:
    as the code is called upon panic, it must not trigger deadlocks or yet
    another Rust panic to fail panic handling.
The careful design of the \projname{} kernel crate frees the cleanup code and
    \texttt{\small drop} handlers of locks and panic-triggering code.
Kernel functions invoked by such code may still hold locks internally, but they
    are self-contained and do not propagate to \projname{} (deadlocks in kernel
    functions is out of the scope of \projname{}).
\projname{} does not execute user-supplied \texttt{\small drop} handlers
    upon panic, as they are not guaranteed to be safe under panic handling
    context.
}

\projname{} implements a crash-stop failure model---a panicked
    extension is removed from the kernel.
Any used maps and other \projname{} extensions sharing the maps
    will also be removed recursively.
This prevents extensions
    sharing the maps from running in a potentially inconsistent
    state---exception handling in \projname{} already ensures the kernel is
    left in a good state.

\vspace{-4pt}
\subsection{Kernel stack safety}
\label{principle:stack}
\vspace{-3pt}

Kernel extensions should never overflow the kernel stack.
Unlike userspace stacks which grow on demand with a large maximum size,
    the stack in kernel space has a fixed size (4 pages on x86-64).
The eBPF verifier checks stack safety by calculating stack size
    via symbolic execution.
However, it is reported that stack safety is broken in eBPF due to the difficulties
    of statically analyzing indirect tail calls~\cite{ebpf-stackoverflow}
    and uncontrolled program nestings~\cite{chintamaneni-ebpf24}.\footnote{Rex currently does not
    support program nesting (same as eBPF)}

Our insight is that stack safety can be enforced
    at compile time to avoid runtime overhead if the extension program has no indirect or
    recursive calls, as 
    the stack usage can be statically computed.
Otherwise,
    it is easy to check stack safety at
    runtime. 
\projname{}, therefore, takes a hybrid approach and selects between static and dynamic checks based on the situation.

\para{Static check.}
The static check is done by a \projname{}-specific compiler pass (\S\ref{sec:impl}).
\new{
If the extension has no indirect or recursive calls,
    its total stack usage can be calculated by traversing its global static
    callgraph and sum up the size of each call frame.
We turn on fat LTO and use a single Rust codegen unit~\cite{rustc-codegen} for
    \projname{} programs to ensure the compiler always has a global view across
    all translation units.
}

\para{Runtime check.}
For extensions with indirect or recursive calls, it is hard to calculate the
    stack usage from the callgraph due to the
    presence of unknown edges (indirect calls) and cycles (recursive calls).
Under these cases, \projname{} performs runtime checks.
The \projname{} compiler pass first ensures each function in the program takes
    less than one page (4K) of stack.
This is more relaxed than the frame size warning threshold (2K) in Linux
    and ensures enough stack to handle Rust panics.
Before each function call in the extension, the compiler inserts a
    call to the \texttt{\small rex\_check\_stack} function from the kernel crate to check the
    current stack usage: if the stack usage exceeds the
    threshold, it will trigger a Rust panic and terminate the
    program safely (\S\ref{principle:eh}).

To manage stack usage of Rex extensions effectively, \projname{}
    implements a dedicated kernel stack for each
    extension.
The dedicated stacks are allocated per-CPU and virtually mapped at
    kernel boot time with a size of eight pages.
Before executing a Rex extension,
    the dispatcher (Figure~\ref{fig:eh-overview}) saves the stack
    pointer of the current context, and
    then sets the stack and the frame pointer
    (already saved with other callee-saved registers) to the
    top of the dedicated stack.
When the extension exits, the original stack and frame pointers are restored.


\projname{} sets the stack usage threshold to be four pages for extension
    code; it reserves the next four pages with following considerations:
(1) helper functions are not visible at compile time but they
    also account for stack usage during execution;
    we use four pages as the de facto stack size used by the kernel itself, and
(2) since stack usage of each function is limited to
    one page of stack, in the worse case, the remaining stack space is at least
    three pages when \texttt{\small rex\_check\_stack} triggers a Rust panic.
\new{
Since the panic handler is implemented in the kernel crate and does not change
    with programs, this worse-case guarantee empirically leaves enough space for
    panic handling and stack unwinding.
}
\projname{}'s dynamic approach achieves stronger stack safety than that of
    eBPF.

\vspace{-4pt}
\subsection{Termination}
\label{principle:termination}
\vspace{-3pt}

Termination is an important property of kernel extensions.
In eBPF, an extension with a back edge or exceeds the instruction limit will be
    rejected, regardless whether it eventually terminates. 
KFlex~\cite{dwivedi-sosp24} lifts the back edge restriction by inserting
    cancellation points in eBPF bytecode on all back edges during
    verification, which triggers termination at runtime.
However, back edge analysis is non-trivial outside
    eBPF bytecode and is unreliable for general Rust programs.
\projname{} employs a runtime that
    interrupts and terminates extensions that run for too long.
\projname{} limits the run time of extensions by
    leveraging kernel timers as watchdogs.
Rex builds the runtime on the high resolution timer
    (\texttt{\small hrtimer}) subsystem in Linux~\cite{linux-hrtimer}.
Since \texttt{\small hrtimer} callbacks execute in hardware timer interrupts,
    they are capable of interrupting the contexts in which most extensions
    execute (soft interrupts and task
    context~\cite{elce-16-chaiken}).
Since hardware timer interrupts are periodically raised
    by the processor, regardless whether an \texttt{\small hrtimer} is present,
    executing timer callbacks in this existing hardware timer interrupts adds
    no extra interrupt or context switch, keeping the
    watchdog overhead minimal.

Rex sets one timer for each CPU to
    avoid inter-core communication, in contrast to using a
    single, global timer to handle programs from all CPUs.
Each timer only needs to monitor extensions running on the core.
Rex arms the timers at kernel boot time, which are triggered periodically
    with a constant timeout, and re-armed each time after
    firing.\footnote{\new{Disarming the timer when no extension is running saves
    CPU cycles, but
    incurs high
    overhead due to timer setup on the hot path of extension execution,
    especially for frequently triggered extensions (e.g., XDP
    extensions~\cite{cilium-docs})}}

\projname{} implements its watchdog logic in the timer handlers.
When a timer fires, its handler
    suspends any soft interrupt or task context, and saves its
    registers.
The handler then checks the current CPU on whether the termination timeout
    of the \projname{} extension in the stopped context has been reached.
This is done by comparing the extension start time (stored as a per-CPU state as
    shown in Figure~\ref{fig:eh-overview}) with the current time.
If the extension exceeds the threshold, the timer handler overwrites the
    saved instruction pointer register to the panic handler (\S\ref{principle:eh}).
After returning from the timer interrupt, the extension executes its
    panic handler, which cleans up kernel resources and gracefully exits.
\projname{} sets both the timer period and runtime threshold to
    the default RCU CPU stall timeout (Rex
    extensions run in RCU locks as they use eBPF hook points). 

\projname{} defers termination when the extension is
    running kernel helper functions to avoid disrupting the kernel's internal resource bookkeeping;
    it also does not terminate an extension if it
    is in the panic handler. 
\projname{} uses a per-CPU tristate flag to track the state of an
    extension: 
    (1) executing extension code, (2) executing kernel helpers or
    panic handlers, and (3) termination requested.
A helper call changes the state from 1 to 2.
When executing the timer handler, if the flag is at state 2,
    the termination handler modifies it to state 3 without
    changing the instruction pointer.
When a helper returns, if the flag is at state 3,
    the panic handler is called to gracefully exit. 

\new{
A corner case of this design is deadlock. Since spinlock acquisition in
    \projname{} is implemented by a kernel helper function, a deadlocked program
    will never return from the helper, and therefore will never be
    terminated properly.
\projname{} follows eBPF's solution toward deadlocks, where a program can only take one lock at a time.
This is achieved by using a per-CPU variable to track whether the program
    currently holds a lock---a program trying to acquire a second lock will
    trigger a Rust panic.
We note that if the ability of holding multiple locks at the same time is
    desired, the kernel spinlock logic can be modified to check the termination
    state of \projname{} programs during spinning and terminate a deadlocked
    program accordingly.
}
\para{Limitation.}
\projname{} uses
    hard interrupts, and thus cannot
    interrupt extensions that are already executing in hard or
    non-maskable interrupts~\cite{elce-16-chaiken} (e.g., hardware perf-event programs).
Such extensions are not targeted by Rex, as they are supposed to be small, simple, and
less likely to encounter the \gap{}.
Note that Rex extensions and eBPF extensions are not mutually exclusive and
    can co-exist.

Moreover, the termination of a timed-out \projname{} extension can be delayed
    if the extension is already interrupted by another event when the timer
    triggers (the registers will not be available to the timer
    handler). \projname{} needs to wait for a triggering of the timer that
    directly interrupts the extension.



%% file: impl.tex
\vspace{-8pt}
\section{Implementation}
\label{sec:impl}
\vspace{-5pt}

\begin{figure*}[t]
    \includegraphics[width=\linewidth]{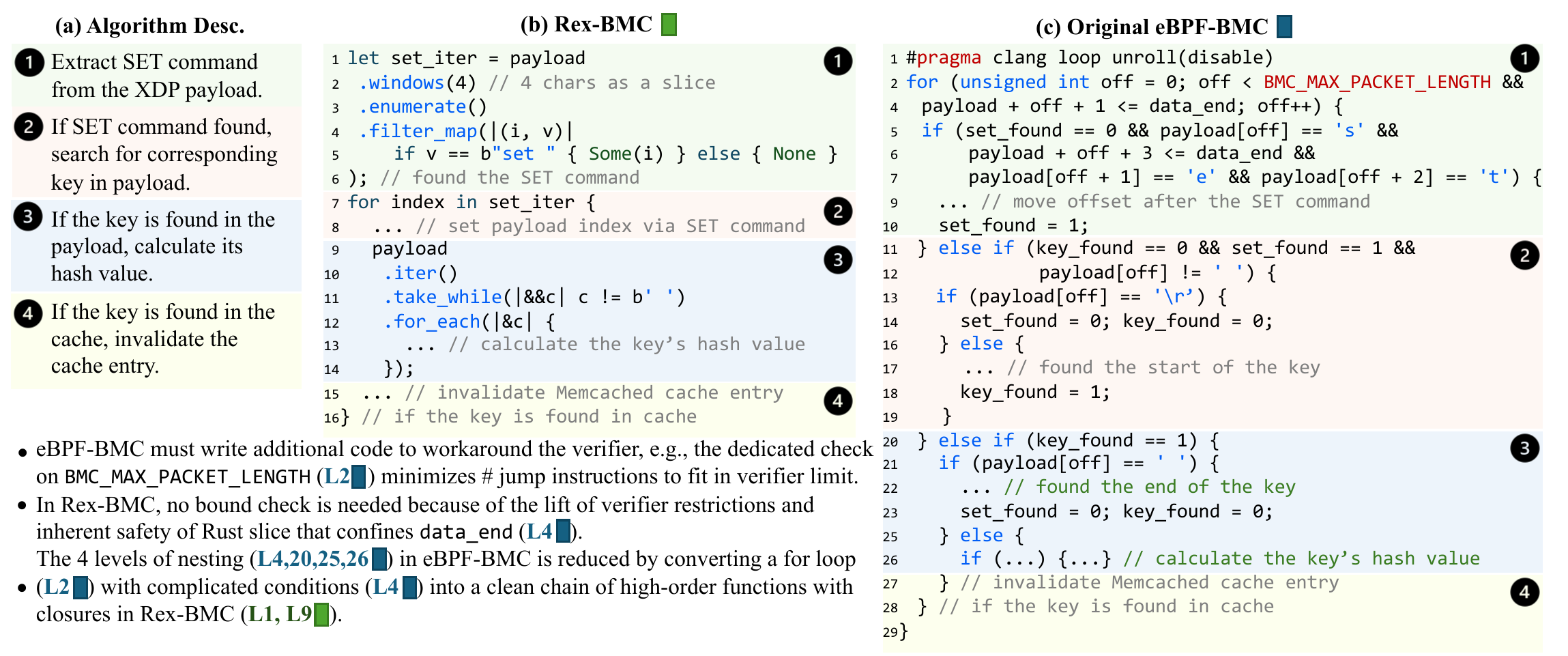}
    \centering
    \vspace{-15pt}
    \caption{Cache invalidation implementation of Rex-BMC and eBPF-BMC; Rex leads to cleaner, simpler code.}
    \vspace{-5pt}
    \label{fig:rust-code}
\end{figure*}

We implement \projname{} on Linux v6.11.
\projname{} currently supports five eBPF program types (tracepoint, kprobe,
    perf-event, XDP, and TC) and shares their in-kernel hookpoints.
Rex only includes helpers for kernel interactions.
helpers introduced due to
    contraints posed by the eBPF verifier
    (e.g., \texttt{\small bpf\_loop}, \texttt{\small bpf\_strtol}, and
    \texttt{\small bpf\_strncmp}) are entirely excluded by \projname{}.

\para{Kernel crate.}
The Rex kernel crate is implemented in 3.5K lines of Rust code, 
    among which 360 lines are unsafe Rust code.
The kernel crate contains the following components:
\begin{packed_itemize}
\item {\it Helper function interface} in Rex is
    implemented on top of eBPF helpers, with
    wrapping code that allows Rex extensions to invoke helpers with safe Rust types.

\item {\it Kernel data-type bindings} are generated for the extension to access kernel
    data types defined in C.
\projname{} uses rust-bindgen~\cite{bindgen} to automatically generate
    kernel bindings and integrates it into the build process of extensions.
\new{
\projname{} programs need to be rebuilt for each kernel they
    target to account for ABI differences in kernel data types.
}

\item {\it Program context} in Rex is wrapped in a Rust struct, which marks
    the context as private and implements getter methods for its
    public fields.
\end{packed_itemize}

\vspace{-5pt}
\para{Kernel support.}
\projname{} implements the extension load code and the runtime in the
    kernel in 2.2K lines of C code on vanilla Linux.
To load an extension, the kernel parses the ELF executable of the extension
    and locates all the \texttt{\small LOAD}
    segments in the executable.
It then allocates new pages and maps the \texttt{\small LOAD} segments into the kernel
    address space based on the size and permissions of the segments.
The load function is responsible for fixups on the program code to resolve
    referenced kernel helpers and eBPF maps.
The \projname{} runtime in the kernel consists of the stack
    unwinding mechanisms (\S\ref{principle:eh}), support for dedicated
    kernel stack (\S\ref{principle:stack}) and
    termination (\S\ref{principle:termination}).

\para{Compiler support.}
\projname{} implements a compiler pass for \projname{}-specific
    compile-time instrumentations on the stack (\S\ref{principle:stack}).
We take advantage of Rust's use of LLVM~\cite{llvm} as its default code
    generation backend and implement the pass in LLVM.
A \projname{}-specific compiler switch is also added to the Rust compiler
    frontend (rustc~\cite{rustc}) to gate the \projname{} compiler pass.

%% file: eval.tex
\vspace{-8pt}
\section{Evaluation}
\label{sec:eval}
\vspace{-5pt}

We evaluate Rex in terms of its usability and performance (with
    both macro and micro benchmarking).

\vspace{-4pt}
\subsection{Usability}
\vspace{-3pt}

Measuring usability is challenging.
We evaluate Rex in two ways: (1) heuristic evaluation on whether it saves workarounds to
    the \gap{}, 
    and (2) our dogfooding experience of using Rex to implement a large, complex extension (BMC~\cite{BMC}).
Overall, we find that Rex enables developers to write simpler and cleaner code.


\para{Eliminating workarounds.}
Since Rex introduces no \gap{}, none of the workarounds in \S\ref{sec:motivation} is
    needed in writing Rex extensions.

\begin{packed_itemize}
\item Rex extensions have no limit on program size and complexity.
    There is no need to artificially refactor extension programs into
    smaller or simpler ones (\S\ref{motivation:restructure}).
\item There is no need to artificially make Rex extensions verifier-friendly (\S\ref{motivation:llvm-codegen}).
In fact, by decoupling static analysis from the kernel,
    Rex can enable new analysis
    (e.g., by allowing compilers to optimize for extra analysis/verification~\cite{wagner:hotos:13}).
\item For the same reason, developers no longer need to tweak code to assist verification (\S\ref{motivation:add-code}).
\item Developers no longer need to manage different verifier bugs across
    kernel versions (\S\ref{motivation:kernel-version}). \new{The Rust compiler
    can have bugs and break safety guarantees, but it is arguably easier to
    upgrade than the kernel for fixes.}
\item Rex enables developers to use rich builtin intrinsics defined by the Rex toolchain
    without reinventing wheels (\S\ref{motivation:wheel}).
\end{packed_itemize}




\para{Case study: \projname{}-BMC}
\label{eval:bmc-case-study}
We rewrite BMC~\cite{BMC} as a Rex kernel extension (\projname{}-BMC), which
    was originally written in eBPF extensions (eBPF-BMC).
Rex-BMC is not a line-by-line translation of eBPF-BMC, 
    because Rex provides more friendly
    programming experience (e.g., no need to split programs due to the verifier limit; see \S\ref{motivation:restructure}).
In this section, we discuss Rex-BMC from the usability perspective
    and measure its performance in \S\ref{eval:macro}.

Our experience shows that Rex enables cleaner and simpler extension code,
    compared to eBPF.
Essentially, Rex enables us to focus on key program logic without the overhead
    of passing the verifier.
For example, we no longer need to divide code into in parts,
    add auxiliary code to help the verifier,
    dealing with tail calls and state transfer, etc.
In addition, we can directly use Rust's builtin language features
    and libraries (e.g., iterators and closures).
As one metric,
    Rex-BMC is written in 326 lines of Rust code.
    In comparison, eBPF-BMC is written in 513 lines of C code (splitting into seven programs).


Figure~\ref{fig:rust-code} compares the code snippets of eBPF-BMC and Rex-BMC that implement
    cache invalidation, respectively, as a qualitative example.
The checks in eBPF-BMC code, required by the eBPF verifier, including these for offset and
    \texttt{\small data\_end} limits, are now being enforced via the inherent language
    features of Rust, such as slices with bound checks in Rex (L2 and L10).
The check on \texttt{\small BMC\_MAX\_PACKET\_LENGTH}, which serves as a constraint to
    minimize the number of jump instructions to circumvent the eBPF verifier,
    is no longer needed.
Other checks for identified SET commands and loops states can be implemented
    with built-in functions and closures in an easy and clean way
    (L4--L6 and L11).
Moreover, with the elimination of program size and complexity limits in \projname{}-BMC,
    developers no longer have to save the computation state in a map
    across tail calls, which leads
    to clearer and more efficient implementation.





    %

Note that the usability benefit does not
    come from the expressiveness difference between Rust and C, but from the
    closing of \gap{} via
    \projname{}.
Evidently, the cleaner code of \projname{}-BMC would fail the
    verifier if it were to be compiled into eBPF (e.g., via Aya~\cite{aya-rs}):
    the compiler is unable to generate verifier-friendly code for convenient
    language features such as slices, and the verifier
    complexity limits will always be an issue.
\projname{} allows us to fully leverage Rust's expressiveness
    without being constrained by verification issues.



\vspace{-4pt}
\subsection{Macro benchmark}
\label{eval:macro}
\vspace{-3pt}

Rex's usability benefits do not come with a performance cost.
We show that Rex extensions deliver comparable performance as eBPF extensions.
\projname{}-BMC achieves a throughput of 1.98M requests per
    second (RPS) on 8 cores, which is slightly higher than eBPF-BMC (1.92M).

%

Our setup consists of a server machine and a client machine.
The server machine runs the \projname{} custom kernel based on Linux v6.11.0 on
    an AMD EPYC 7551P 32-Core processor with 112 GB memory without SMT and
    Turbo.
The client machine runs a vanilla v6.11.0 Linux kernel on an AMD Ryzen 9 9950X
    processor with 96 GB memory.
Both machines are equipped with Mellanox ConnectX-3 Pro 40GbE NICs and are
    connected back-to-back using a single port.



We evaluate the throughput of (1) Memcached which binds
    multiple UDP sockets to the same port~\cite{BMC},
    (2) Memcached with eBPF-BMC,
    and (3) Memcached with \projname{}-BMC.
For each setup, we vary the number of CPU cores for Memcached server and
    NIC IRQs and pin one Memcached thread onto each available core.
We use the same workloads as in BMC~\cite{BMC}, albeit with a smaller
    number of Memcached keys.

\begin{figure}
    \includegraphics[width=0.9\linewidth]{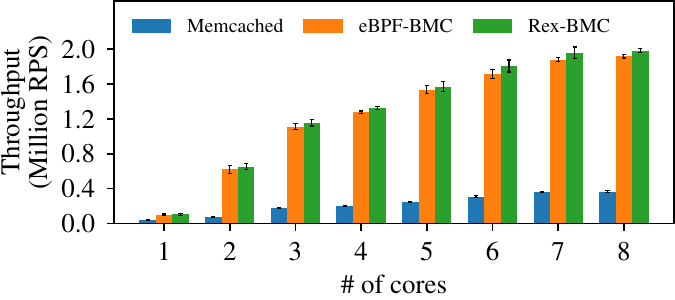}
    \centering
    \vspace{-2pt}
    \caption{Throughput of Memcached, with eBPF-BMC, and with \projname{}-BMC
        under different number of cores.
    }
    \label{fig:eval-bmc}
    \vspace{-5pt}
\end{figure}

Figure~\ref{fig:eval-bmc} shows the throughput of the three setups under
    different numbers of CPU cores.
Memcached processes all requests in userspace with the overhead of the kernel network stack,
    achieving only 37K RPS on a single core and 365K RPS on 8 cores.
Both eBPF-BMC and \projname{}-BMC
    achieve a much higher throughput as they process a large
    fraction of requests at NIC driver level without going through
    the kernel network stack.
With 8 cores, eBPF-BMC and \projname{}-BMC achieve a throughput
    of 1.92M and 1.98M, and a performance benefit of 5.26x and 5.43x, respectively.
The slight performance improvement over eBPF is attributable to the
    elimination of overheads of tail calls and associated state-passing
    via maps, 
    along with optimizations in the rustc frontend and
    x86 backend, 
    despite the overhead of additional runtime checks.

\vspace{-4pt}
\subsection{Micro benchmark}
\vspace{-3pt}

Several of Rex's designs could introduce overheads,
    despite invisible in the \projname{}-BMC evaluation.
We use microbenchmarks specifically designed to stress our design and measure
    overheads.
We show that overheads exist
    in some pessimistic cases, but have negligible impact in real-world scenarios.
All experiments are performed on the same machine that acts as the server in
    the \projname{}-BMC experiments (\S\ref{eval:macro}).

\para{Setup and teardown.}
Entering and exiting a \projname{} program requires \projname{}-specific
    operations 
    (Figure~\ref{fig:eh-overview}).
\projname{}'s use of a dedicated stack requires saving the stack
    pointer and setting the new stack and frame pointer to the dedicated stack
    (and restoring to the saved values after the extension exits).
\projname{} also needs to set up the per-CPU state used by its
    termination mechanism (\S\ref{principle:termination}).
In total, these operations add eight instructions on the
    execution path in \projname{}.
To measure the overhead, we implement an empty extension program in both
    eBPF and
    \projname{} and record their execution time (including the program dispatcher).
As shown in Table~\ref{tab:startup-cleanup}, the measured execution time
    of the empty \projname{} and eBPF
    programs only differ in around a nanosecond on average.


\para{Exception handling}
\label{eval:termsupport}
Rex's safe cleanup for exception handling requires recording allocated
    resource at runtime (\S\ref{principle:eh}), which, compared to eBPF, adds overhead.
We measure the overhead using a program that acquires and then
    immediately releases an eBPF spinlock.
Since the acquired spinlock needs to be released upon Rust
    panics, \projname{}'s cleanup mechanism records it in its per-CPU
    buffer.
Additionally, \projname{} sets up a per-CPU state flag to indicate
    execution of a helper function (\S\ref{principle:termination}).
The program is implemented in both eBPF and \projname{} and the time used to
    acquire and release the spinlocks are measured.
Table~\ref{tab:startup-cleanup} shows that the runtime difference between
    eBPF and \projname{} is roughly 50 nanoseconds.



\begin{table}[t]
    \small
    \centering
    \vspace{-10pt}
    \caption{Time to execute an empty extension program and
        to acquire and release a spinlock in eBPF and \projname{} (nanosecond)}
    \begin{tabular}{ccc}
        \toprule
        \textbf{Extension} & \textbf{Empty prog runtime} & \textbf{Spinlock runtime} \\
        \midrule
        eBPF & 42.1 $\pm$ 4.1 ns & 130.4 $\pm$ 20.3 ns\\
        \projname{} & 42.6 $\pm$ 5.8 ns & 183.1 $\pm$ 27.5 ns\\
        \bottomrule
    \end{tabular}
    \label{tab:startup-cleanup}
\end{table}

\begin{figure}[t]
    \vspace{10pt}
    \centering
    \begin{minipage}{0.23\textwidth}
        \centering
        \includegraphics[width=\textwidth]{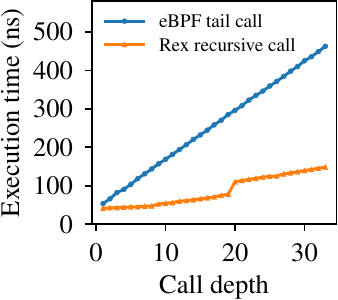}
        \caption{eBPF tail call and \projname{} recursive call time}
        \vspace{-10pt}
        \label{fig:eval-recursion}
    \end{minipage}%
    \hfill
    \begin{minipage}{0.23\textwidth}
        \centering
        \includegraphics[width=\textwidth]{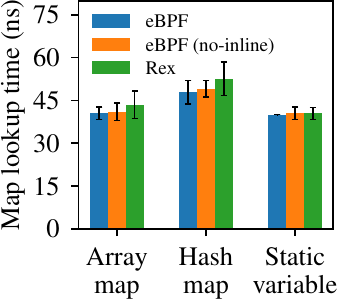}
        \caption{Map lookup time under various setups}
        \vspace{-10pt}
        \label{fig:map-bench}
    \end{minipage}
\end{figure}

\para{Stack check.}
Stack checks are added before
    function calls in \projname{} extensions that contain indirect
    or recursive calls (\S\ref{principle:stack}).
We implement recursive extension programs in both
    eBPF and \projname{} to measure the overhead.
The recursive function calls itself for a controlled number of times.
In \projname{}, we pass the call depth as the argument to the recursive
    function;
since eBPF does not support recursive functions, we use eBPF tail
    calls to implement the logic---since it is inconvenient to pass arguments to
    tail-called programs (\S\ref{motivation:restructure}), we use a static
    variable to set the call depth.
Figure~\ref{fig:eval-recursion} plots execution time of the recursive programs with
    call depths from 1 to 33 (eBPF cannot do more than 33 tail calls).
\projname{} is roughly 3x faster than eBPF.
The overhead of eBPF is due to runtime check on tail-call
    count limit and accessing the static variable,
    which is a map in eBPF 
    (not a register in normal calls).


\para{Map access}
\label{eval:map}
Map access in Rex is expected to have more overhead than in eBPF.
First, \projname{} implements wrapping code to enforce safety
    of helper function calls (\S\ref{sec:impl}).
Moreover, the eBPF JIT compiler inlines the helper function for map lookup
    at load time
    as a performance 
    optimization;
however, inlining is not available in \projname{} (no JIT in Rex).
We measure map lookup time of Rex, compared with
    eBPF with and without inlining, including
    array map, hash map, and static variable.
In eBPF, static variables are
    converted into maps;
we use a static Rust atomic variable in Rex, as the counterpart of a
    static variable map in eBPF.
Figure~\ref{fig:map-bench} shows the lookup time of different
    maps in eBPF and \projname{}, respectively.
We find that
    inlining map lookups in eBPF are $\sim$0.5$ns$
    faster on array maps and $\sim$1.2$ns$ faster on hash maps.
An additional slowdown of 2$ns$--4$ns$ is present in
    \projname{} over non-inlined eBPF, due to the wrapping
    code.
Static variables in eBPF are always accessed via direct load without
    invoking a helper.
Hence, their access latency is almost the same to accessing Rust atomic variables.

%% file: discussion.tex
\vspace{-8pt}
\section{Discussion}
\label{sec:discussion}
\vspace{-5pt}


{\bf Verification without language-verifier gaps.} Rex currently uses
    language features of Rust to ensure safety of kernel extensions.
This approach defers the checking of some safety properties to the runtime
(e.g., termination, integer errors).
It may be possible to minimize the amount of runtime errors by incorporating
    Rust-based verification
    techniques, e.g., ensuring freedom of panics~\cite{verus,verus-sosp24,paniccheck,rvt,kani-rust}.
Certainly, push-button verification techniques that use symbolic execution
    such as PanicCheck~\cite{paniccheck} are likely to
    re-introduce the \gap{}.  We suspect that using
    verification techinques that combine proofs and programming~\cite{coq,dafny,verus-sosp24,verus},
    such as Verus for Rust may allow \projname{} to reduce
    runtime errors \textit{without} the \gap{}.

\para{Dynamic memory allocation.}
eBPF has recently supported dynamic allocation~\cite{Dwivedi-958cf2e273f0}
    that allows extension programs to request memory
    from the kernel using allocation kfuncs~\cite{kfuncs}.
\projname{} currently does not support dynamic memory allocation.
We plan to integrate memory
    allocation~\cite{rust-alloc} of Rust with the eBPF all-context
    allocator~\cite{bpf-mempool-lwn}, granting \projname{} dynamic allocation.
Dynamic allocation enhances programmability of extension programs and opens
    up more advanced use cases~\cite{dwivedi-sosp24}.
It also makes more components of the Rust standard library available, notably
    the collection and smart pointer types with automatic memory management.

\para{Kernel crate maintenance.}
The \projname{} kernel crate inevitably needs to use unsafe Rust, as it
    directly interacts with kernel functions and variables. 
As a principle, unsafe Rust code must not be used for escaping
    safety checks but only when it is the last resort
    (mostly for foreign function interface, FFI).
This keeps the scope of unsafe Rust at its minimum---the \projname{}
    kernel crate only leverages unsafe Rust necessary for FFI
    interaction and contributes to about 10\% (360 lines) of kernel crate code.
As unsafe code is isolated from extension programs and managed at a central
    location by trusted maintainers, we are not particularly concerned about
    its maintainability.

%% file: related.tex
\vspace{-8pt}
\section{Related Work}
\label{sec:related}
\vspace{-5pt}

{\bf Improving eBPF.} eBPF has evolved from simple
    use cases like packet filtering~\cite{pf,bsdpf}
    into a general-purpose kernel extension language and programming framework
    that enables many innovative projects~\cite{BMC,Electrode,DINT,Hoiland-Jorgensen:conext:2018,
    Zhong:osdi:2022,ghost-scheduler-lpc,ebpf-mm,
    ghost-scheduler,lpc-24-bpfmm,fetchbpf,sched-ext}.
Recent work is making active progress to improve the correctness and security
    of the eBPF infrastructure, including
    fuzzing and bug finding~\cite{hung2023brf,hao-eurosys,hao-osdi,ebpf-fuzzing},
    formal verification~\cite{ebpf-jit-formal,proof-carrying-verifier,Wang:2014,hari-cav-verification},
    sandboxing~\cite{sandbpf,beebox-security24,safebpf-thomas},
    and integrating with hardware protection mechanisms~\cite{hive-ebpf-sandbox,lu2024moat}.
eBPF's design, which relies on an in-kernel static verifier for extension safety,
    inevitably creates the \gap{} (\S\ref{sec:motivation}).
In contrast, Rex provides an alternative to develop and maintain large, complex kernel extensions directly with
    high-level language safety, avoiding the \gap{}.




\para{Other frameworks.} The idea of building safe OS components using
    safe languages was proposed by SPIN~\cite{spin}
    and revisited by Singularity~\cite{singularity}, Tock~\cite{tockos}, and a few recent
    discussions~\cite{Miller-hotos19,untenableVerification,Burtsev-hotos23}.
However, adopting them in practice is challenging as they are based on
    clean-slate OS designs.
Rex develops a practical kernel extension framework for Linux, taking the
    opportunity of recent support of Rust as a safe language for OS code.
It addresses the key challenges of enforcing safe code only, interfacing
    with unsafe C code, and providing safety guarantees beyond language-based safety.

KFlex~\cite{dwivedi-sosp24} is a recent kernel extension framework built on top of eBPF.
KFlex aims to improve the flexibility of eBPF to let developers express diverse functionality in extensions.
It employs an efficient runtime by co-designing it with the existing eBPF verifier: (1) its Software Fault Isolation (SFI)
    elides checks already done by the verifier for efficiency,
    and (2) its termination mechanism uses the verifier to
    statically compute the kernel resources acquired by the extension.
Rex made the same design choice as KFlex to use a lightweight runtime for safety properties that are hard to
    check statically.
Unlike KFlex, which is co-designed with the eBPF verifier,
    Rex eliminates the verifier to close the \gap{}.

BCF~\cite{lpc-24-bcf-lazy-abstraction-proof} is a recent proposal to enhance eBPF's in-kernel verification
    with help from user space, asking for proof when the
    verifier fails to reason about certain program properties.
The idea echoes proof-carrying code~\cite{necula-pcc} which asks a program to attach a formal proof
    that its code obeys the safety policy.
BCF leverages the eBPF verifier's range analysis and symbolic execution for proof
    generation but still requires developers to specify safety conditions to aid the generation.
Its uses of the verifier still lead to the \gap{}.

\para{Rust for OS kernels.} Rust has been embraced
    by modern OSes~\cite{rust-for-linux-doc,rust-for-windows} as
    practical language which leads to safer code.
Recent work shows the promises to build new OS kernels using Rust~\cite{redleaf,theseus,tockos,Burtsev-hotos23}.
We claim no novelty of using Rust as a language.
In fact, a safe language alone does not lead to system safety,
    as exemplified by Rust kernel modules~\cite{rust-module-dev-quit-lwn}.
Rex shows an example of how to build upon language-based safety
    to enable and enforce safe kernel extension programs.



%% file: conclusion.tex
\vspace{-8pt}
\section{Conclusion}
\label{sec:conclusion}
\vspace{-5pt}

We build Rex, a new kernel extension framework that closes the \gap{}. 
We believe that closing the gap is essential to  
    programming experience and maintainability of kernel extensions, especially
    those that embody large, complex programs for advanced features.
Rex provides a solution that allows kernel extensions to be developed and maintained
    in a high-level language, while providing desired safety guarantees 
    as the existing framework like eBPF.

\vspace{-8pt}
\section*{Acknowledgement}
\vspace{-5pt}

This work was funded in part by NSF CNS-1956007, NSF
CNS-2236966, and an IBM-Illinois Discovery Accelerator Institute (IIDAI) grant.
We used Rex to develop Machine Problems (MPs) for CS 423 (Operating System Design)
at the University of Illinois Urbana-Champaign in Fall 2022, 2023, and 2024.
We thank students in CS 423 for being the beta users of Rex, which 
    provides valuable feedback for us to improve the usability of Rex.
We thank Minh Phan, Manvik Nanda, and Quan Hao Ng for their participation
    in the project.
We also thank Jiyuan Zhang, Di Jin, Hao Lin, James Bottomley, and Darko Marinov for their
    feedback and discussion.

